\documentclass[12pt,preprint]{aastex}

\def\ffrac#1#2{{\textstyle\frac{#1}{#2}}} \def\C{{\cal C}}
\def\Ccut{{\cal C}_{\rm cut}}
\def\zb{{\bar z}}
\def\Qb{{\bar Q}}
\def\zetab{{\bar \zeta}}

\def\ts{{t_{\rm s}}}
\def\det{{\rm det}}
\def\Re{{\rm Re}}
\def\poles{{\rm poles}}
\def\images{{\rm images}}
\def\Residue{{\rm Residue}}

\def\ds{\displaystyle}

\def\pd#1#2#3{{\ds \partial #1 \over \ds \partial #2}\Bigl|_#3}
\def\spose#1{\hbox to 0pt{#1\hss}}
\def\lta{\mathrel{\spose{\lower 3pt\hbox{$\sim$}} \raise
2.0pt\hbox{$<$}}}
\def\gta{\mathrel{\spose{\lower 3pt\hbox{$\sim$}} \raise
2.0pt\hbox{$>$}}}

\begin{document}
\title{Lensing Properties of Scale-Free Galaxies}
\author{C. Hunter}
\affil{Department of Mathematics, Florida State University,
Tallahassee, Florida 32306-4510, USA \\ 
hunter@math.fsu.edu} 
\author{N. W. Evans}
\affil{Theoretical Physics, Department of Physics, 1 Keble Road, 
Oxford, OX1 3NP, UK \\
w.evans1@physics.oxford.ac.uk}


\begin{abstract}
\noindent 
The multiple images of lensed quasars provide evidence on the mass
distribution of the lensing galaxy. The lensing invariants are
constructed from the positions of the images, their parities and their
fluxes. They depend only on the structure of the lensing
potential. The simplest is the {\it magnification invariant}, which is
the sum of the signed magnifications of the images. Higher order {\it
configuration invariants} are the sums of products of the signed
magnifications with positive or negative powers of the position
coordinates of the images.

We consider the case of the four and five image systems produced by
elliptical power-law galaxies with $\psi \propto (x^2 + y^2
q^{-2})^{\beta/2}$. This paper provides simple contour integrals for
evaluating all their lensing invariants. For practical evaluation,
this offers considerable advantages over the algebraic methods used
previously. The magnification invariant is exactly $B = 2/(2-\beta)$
for the special cases $\beta =0, 1$ and $4/3$; for other values of
$\beta$, this remains an approximation, but an excellent one at small
source offset. Similarly, the sums of the first and second powers of
the image positions (or their reciprocals), when weighted with the
signed magnifications, are just proportional to the same powers of the
source offset, with a constant of proportionality $B$. To illustrate
the power of the contour integral method, we calculate full expansions
in the source offset for all lensing invariants in the presence of
arbitrary external shear.  As an example, we use the elliptical
power-law galaxies to fit to the data on the four images of the
Einstein Cross (G2237+030). The lensing invariants play a role by
reducing the dimensionality of the parameter space in which the
$\chi^2$ minimisation proceeds with consequent gains in accuracy and
speed.

\end{abstract}


\section{Introduction}

\noindent
There are now some sixty or so multiply imaged quasars known. Most of
these exhibit double or quadruple images, although some triples and
rings are also known (see Warren et al. 1999, Ratnatunga, Griffiths,
Ostrander 1999, Wisotzki et al. 1999 and Tonry \& Kochanek 1999 for
some recent examples, as well as Pospieszalka et al. 1999 for details
of the gravitational lensing database which maintains a list of
candidates). This paper is mainly concerned with the sixteen or so
quadruple systems. In propitious circumstances, the observables of
such systems are the location of the lens, the positions of the four
images, their parities and their fluxes. Lensing invariants are
constructed directly from the lensing observables and remain
invariant against changes in the source or lens configuration
(provided a caustic is not crossed). A classical problem in the
theory of gravitational lensing is the construction of a lens model
that reproduces the observed properties of the images. One of the
advantages of lensing invariants is that they provide simple tests by
which we can determine quickly whether a given set of images can be
produced by a particular lens.

Witt \& Mao's (1995) paper provides the first example of a lensing
invariant. They considered a binary lens of two point masses, which
has either three or five images according to whether the source lies
outside or inside the caustic (e.g., Schneider, Ehlers \& Falco 1992,
section 8.3). Their main result is that the sum of the signed
magnifications of the five images is unity when the source lies
within the caustic, namely
\begin{equation}
\label{eq:wittmao}
\sum_{i=1}^5 \mu_i p_i = 1.
\end{equation}
Here, $\mu_i$ is the absolute magnification of the image, while $p_i$
is the parity. This result holds good irrespective of the position of
the source, provided it remains within the central caustic.
Subsequently, Rhie (1997) showed that the sum of the signed
magnifications of the images of the $N$ point mass lens is also equal
to unity, provided the source location yields the maximum number of
$N^2 +1$ images. These are remarkably simple results, bearing in mind
that the lens equation that provides the locations of the images must
be solved numerically. Further examples of lensing invariants were
found by Dalal (1998) and Dalal \& Rabin (2000).

Witt \& Mao's (2000) paper provides a study of lensing potentials
stratified on similar concentric ellipses and falling off like a
power-law with index $\beta$ suitable for modelling quadruple lenses.
They showed that for the point mass ($\beta =0$) and isothermal
($\beta =1$) cases, there is an invariant 
\begin{equation}
\label{eq:wittmaoagain}
\sum_{i=1}^4 \mu_i p_i = {2 \over 2 -\beta}=B. 
\end{equation} 
They introduced higher-order moments and showed that they too had
simple results in the point mass and isothermal cases 
\begin{equation}
\label{eq:wittmaomore}
\sum_{i=1}^4 \mu_i p_i x_i= B \xi,
\qquad \sum_{i=1}^4 \mu_i p_i y_i= B\eta. 
\end{equation} 
Here, $(x_i,y_i$) are the coordinates of the $i$th image, whereas
$(\xi,\eta$) is the position of the source. Witt \& Mao (2000) argued
on the basis of numerical experimentation that
(\ref{eq:wittmaoagain}) and (\ref{eq:wittmaomore}) remain good
approximations for all other values of $\beta$. They suggested the
use of interpolation to extend the result to all the power-law
potentials. Witt \& Mao's (2000) paper is especially important to us
here, as our results extend the work that they began.

In \S 2, we exploit contour integral methods to provide a new way of
finding the invariants of a gravitational lens system. The idea is
very simple. The images are the roots of the lens equation. So, by a
judicious choice, the observables at each image can be made to equal
the residue of a complex integrand. The lensing invariants, which are
sums over the images, then become equal to a contour integral by
Cauchy's theorem. Although our method is more general, this paper
concentrates on applications to the elliptic power-law potentials. In
\S3 we evaluate invariants for the case in which $B=2/(2-\beta)$ is a
positive integer. Our complex integrand is then single-valued. The
simplest cases are the point mass ($\beta =0$, $B=1$) and the
isothermal ($\beta =1$, $B=2$) lenses noted by Witt \& Mao (2000).
The invariants for $B=3$ include the contributions from a weak fifth
image, while those for integer $B>3$ include small contributions from
spurious images, which are non-physical solutions of the lens
equation.

In \S 4, we progress to the case of non-integer $B$ when the
contributions from a branch cut must be added to the earlier results.
Remarkably this gives us exact expressions for the lensing invariants
for the four bright images for all $B$, because the added
contributions cancel out the effects of the fifth and the spurious
images. Thus, our contour integral method provides not only a simpler
but also a more powerful way to determine the lensing invariants than
the earlier algebraic methods. Astrophysical applications of the
invariants are presented in \S 5. They are useful because they can
short-cut the modelling process, telling us quickly whether an assumed
lensing potential (in this case the elliptic power-law potentials) can
reproduce the data. We pay particular attention to the Einstein Cross
gravitational lens system (G2237+0305) as our example. Finally, \S 6
discusses our conclusions and presents plans for future work.

\section{The Contour Integral Method}

\subsection{The Lens Equation}

At outset, we shall assume that the lensing potential is stratified
on similar concentric ellipses
\begin{eqnarray}
\label{eq:lensing_pot}
\psi =\left\{ \begin{array}{ll} {\displaystyle A\over \displaystyle
\beta} (x^2 + y^2q^{-2})^{\beta/2} & \mbox{if $0 < \beta < 2$}, \\
\null&\null\\ {\displaystyle A\over \displaystyle 2} \log (x^2 +
y^2q^{-2} ) & \mbox{ if $\beta =0$}. \end{array} \right.
\end{eqnarray}
This family of models was introduced into gravitational lensing by
Blandford \& Kochanek (1987) and subsequently studied by others
(e.g., Kassiola \& Kovner 1993; Witt 1996; Witt \& Mao 1997, 2000;
Evans \& Wilkinson 1998). Here, $q$ is the axis ratio of the
projected equipotentials and is chosen to satisfy $0 <q \le 1$
without loss of generality. The parameter $\beta$ controls the radial
profile of the potential while $A$ measures the depth of the well.
The point mass (Schwarzschild lens) has $\beta =0$ and $q=1$, while
the isothermal sphere has $\beta =1$ and $q=1$. The gravitational
convergence (or projected density) is
\begin{equation}
\kappa = {A\over 2q^2}{[ 1+ q^2(\beta-1)]x^2 + [1 +
q^{-2}(\beta-1)]y^2 \over
(x^2 + y^2 q^{-2})^{2 - \beta/2}}.
\end{equation}
This is positive definite provided $\beta > 1- q^2$.
Three-dimensional analogues of these models are already familiar in
galactic dynamics as the power-law models (Evans 1993, 1994). They are
the only flattened and reasonably realistic galaxy models known, for
which the self-gravitation equations can be solved to find simple
two-integral distribution functions. For example, they have been used
in modelling the nearby elliptical galaxy M32 (van der Marel et al.
1994), the inner parts of the Galactic bulge (Evans \& de Zeeuw 1994),
as well as the dark halo of our own Galaxy in the interpretation of
the microlensing results (Alcock et al. 1997).

For the geometrically thin lens with projected potential
(\ref{eq:lensing_pot}), the paths of photons are given by (e.g.,
Schneider, Ehlers \& Falco 1992, section 5.1) 
\begin{eqnarray}
\label{eq:original}
\xi = x + \gamma_1 x + \gamma_2 y - {\displaystyle Ax \over
\displaystyle (x^2 + y^2 q^{-2})^{1-\beta/2}},\qquad\qquad
\eta = y + \gamma_2 x - \gamma_1 y - {\displaystyle Ay q^{-2}\over
\displaystyle (x^2 + y^2 q^{-2})^{1-\beta/2}}, 
\end{eqnarray}
where ($\xi, \eta$) are Cartesian coordinates of the source. Here,
$\gamma_1$ and $\gamma_2$ allow for a constant external shear in an
arbitrary direction. The external shear may be produced by the
gravity field of a cluster or by the effects of nearby galaxies and
usually lies between $0 < |\gamma| < 0.3$ (Keeton, Kochanek \& Seljak
1997). We shall always assume that the lens is not circularly
symmetric (that is, either $q\neq 1$ or $\gamma_1 \neq0$ or $\gamma_2
\neq0$). The lensing properties of circularly symmetric lenses are
qualitatively different from those of more realistic flattened
lenses. For example, the tangential caustic degenerates to a point in
the case of circular symmetry (e.g., Schneider et al. 1992, section
8.1). So, circularly symmetric lenses can give misleading results and
they are not widely used in modelling.

A formulation in terms of complex numbers has been used before to
ease calculations in lensing theory (e.g., Bourassa, Kantowski \&
Norton 1973, Bourassa \& Kantowski 1975, Witt 1990). However, we
shall find it helpful to use somewhat different definitions, namely 
\begin{equation}
\zeta = \xi + {\rm i} q \eta, \qquad\qquad z = x + {\rm i}y/q. 
\end{equation} 
The lens equation (\ref{eq:original}) becomes 
\begin{equation}
\zeta = \ffrac{1}{2} [1+q^2 + \gamma_1(1-q^2)] z + \ffrac{1}{2}
[1-q^2 + \gamma_1(1+q^2) + 2{\rm i} q \gamma_2] \zb - {\displaystyle Az
\over \displaystyle (z\zb)^{1-\beta/2}}. 
\end{equation}
It is convenient to introduce a variable $t$ such that: 
\begin{equation}
\label{eq:deft}
t = {A \over (z\zb)^{1/B}}, \qquad\qquad B = {2\over 2- \beta},
\end{equation}
where the bar represents complex conjugation, and henceforth to set
the constant $A$ equal to unity. The dependence of our results on $A$
can be recovered by dividing all powers of $z$ and $\zeta$ and their
conjugates by the same powers of $A^{B/2}$. Then, the lens equation
can be written in matrix form as
\begin{displaymath}
\left( \begin{array}{l}
\zeta \\
\zetab \\ \end{array} \right)
= \left( \begin{array}{lr}
P & Q \\
\Qb & P \\
\end{array} \right)
\left( \begin{array}{l}
z \\
\zb \\ \end{array} \right),
\end{displaymath}
with
\begin{equation}
P =\ffrac{1}{2} [ 1+ q^2 + \gamma_1(1-q^2)] - t=P_0-t,\qquad Q
=\ffrac{1}{2} [ 1- q^2 + \gamma_1(1+q^2) + 2{\rm i}q\gamma_2 ],
\end{equation}
from which it follows, by the standard algorithms of matrix
inversion, that
\begin{equation}
\label{eq:zzetarel}
\left( \begin{array}{l}
z \\
\zb \\ \end{array} \right)
= {1\over P^2 - |Q|^2}\left( \begin{array}{lr} P & -Q \\ -\Qb & P \\
\end{array} \right)
\left( \begin{array}{l}
\zeta \\
\zetab \\ \end{array} \right).
\end{equation}
By forming $t^B = 1/(z\zb)$, we see that the solutions of the lens
equation also satisfy the equation
\begin{equation}
\label{eq:defnk}
K(t; \zeta, \zetab) \doteq t^B (P\zeta - Q \zetab )(P\zetab -\Qb
\zeta) - \Biggl[ P^2 - |Q|^2\Biggr]^2 =0. \end{equation}
Let us designate $K(t, \zeta, \zetab) =0$ as {\it the imaging
equation}. Its real and positive solutions provide the possible image
positions $t$ (or equivalently $z$), given the source location $\zeta$
and its conjugate $\zetab$. The imaging equation must have at least as
many solutions as the original lens equation (\ref{eq:original}). It
may have more; that is, there may be solutions of (\ref{eq:defnk})
that do not correspond to true images and which we will call {\it
spurious roots}.

For example, when $B$ is an integer $N \le 2$, then the imaging
equation is a quartic with four roots each of which corresponds to one
of the images of a quadruple lens. When $B = N >2$, then the imaging
equation is a polynomial of degree $N+2$. In the case $B=3$, the roots
all correspond to the five images of a quintuple lens. However, when
$B>3$, there are always roots of the imaging equation that are
spurious, as there are never more than five images. This is shown in
Appendix A.

\subsection{The Magnification Invariant}

Mathematically speaking, the lens equation defines a map from the
lens plane to the source plane (see Schneider et al. 1992, chapter
5). The Jacobian matrix of this mapping is denoted by $J$, where 
\begin{displaymath}
J = \left( \begin{array}{lr}
\pd{\zeta}{z}{\zb} &
\pd{\zeta}{\zb}{z} \\
\pd{\zetab}{z}{\zb} &
\pd{\zetab}{\zb}{z}
\end{array} \right)\qquad\qquad
J^{-1} = \left( \begin{array}{lr}
\pd{z}{\zeta}{\zetab}
& \pd{z}{\zetab}{\zeta} \\
\pd{\zb}{\zeta}{\zetab} &
\pd{\zb}{\zetab}{\zeta}
\end{array} \right).
\end{displaymath}
With our choice of source and image variables, the reciprocal of the
determinant of $J$ corresponds physically to the signed magnification
of an image via
\begin{equation}
\label{eq:magperturb}
{q^2\over \det J}\Biggl|_{x_i,y_i} = \mu_i p_i, \end{equation} 
where $\mu_i$ is the absolute value of the magnification and $p_i$ is
the parity of the image located at $(x_i, y_i)$. For certain
positions, $\det J$ may vanish and the magnification is infinite.
These are the critical points and lines. The caustics are the images
of the critical points and curves under the lens mapping
(\ref{eq:original}).

By differentiating the expression for $t$, we obtain: 
\begin{eqnarray}
-{B\over t^{1+B}}\pd{t}{\zeta}{\zetab} & = & \pd{z}{\zeta}{\zetab}
\zb + \pd{\zb}{\zeta}{\zetab} z, \\
-{B\over t^{1+B}}\pd{t}{\zetab}{\zeta} & = & \pd{z}{\zetab}{\zeta}
\zb + \pd{\zb}{\zetab}{\zeta} z.
\end{eqnarray}
Recalling the properties of the Jacobian matrix $J$ and its inverse
$J^{-1}$, it follows that
\begin{eqnarray}
\label{eq:one}
\pd{\zeta}{z}{\zb}\pd{t}{\zeta}{\zetab} + \pd{\zetab}{z}{\zb}
\pd{t}{\zetab}{\zeta} & = & -{t^{1+B} \over B} \zb, \\ \label{eq:two}
\pd{\zeta}{\zb}{z} \pd{t}{\zeta}{\zetab} + \pd{\zetab}{\zb}{z}
\pd{t}{\zetab}{\zeta} & = & -{t^{1+B} \over B} z. 
\end{eqnarray} 
By multiplying (\ref{eq:one}) by $\zetab_\zb$ and (\ref{eq:two}) by
$\zetab_z$ and subtracting, we construct the determinant of the
Jacobian matrix and so establish
\begin{equation}
\label{eq:deter}
\det J \, \pd{t}{\zeta}{\zetab} = -{t^{1+B} \over B} \Biggl(
\pd{\zetab}{\zb}{z} \zb - \pd{\zetab}{z}{\zb} z \Biggr).
\end{equation}
However, equation (\ref{eq:defnk}) provides us with the requirement
that $K(t; \zeta, \zetab) =0$ and so, using a standard formula in the
theory of partial differentiation, we have 
\begin{equation}
\label{eq:detera}
{1\over \det J} = { B \over t^{1+B} \Bigl( \pd{\zetab}{\zb}{z} \zb
- \pd{\zetab}{z}{\zb} z \Bigr)}\, {\pd{K}{\zeta}{{\zetab,t}} \over
\pd{K}{t}{{\zeta,\zetab}}}.
\end{equation}
Now comes a remarkable simplification. For the ellipsoidally
stratified potentials, we find by direct evaluation that %
\begin{equation}
\pd{\zetab}{\zb}{z} \zb - \pd{\zetab}{z}{\zb} z = { (P^2 +
|Q|^2)\zetab -2P \Qb \zeta \over P^2 - |Q|^2}. 
\end{equation} 
A similar factor occurs on partially differentiating $K(t; \zeta,
\zetab)$; in other words
\begin{equation}
\pd{K}{\zeta}{{\zetab,t}} = t^B\,\Biggl[(P^2 + |Q|^2) \zetab - 2P \Qb
\zeta \Biggr].
\end{equation}
So, equation (\ref{eq:detera}) simplifies to 
\begin{equation}
\label{eq:defmag}
{1\over \det J} = { B \over t }\, {P^2 - |Q|^2 \over
\pd{K}{t}{{\zeta,\zetab}}}.
\end{equation}
Let us recall that the images correspond to simple zeros of $K(t;
\zeta, \zetab)$ and the residue of $1/K$ at a simple zero is just $1/
K_t$. This gives us a contour integral representation of the {\it
magnification invariant}, which is defined as the sum of the signed
magnifications of the images
\begin{equation}
\label{eq:theorem}
\sum_{\images} \mu_i p_i = {q^2 B \over 2 \pi {\rm i}} \oint_\C {dt\over t}
{ P^2 - |Q|^2 \over K(t; \zeta, \zetab)}, \end{equation} 
where $\C$ is a contour in the complex lens plane enclosing only the
simple poles corresponding to visible images. It is an invariant, as
it is a pure number which depends only on the structure of the
lensing potential or, equivalently, the mass distribution.

\subsection{The Configuration Invariants}

The higher order invariants are sums of powers of the coordinates of
the images weighted with their signed magnifications. We shall call
them the {\it configuration invariants}.  From the lens equation
(\ref{eq:original}), we deduce that
\begin{eqnarray}
x(t; \zeta, \zetab) &=& { q^2[(1 - \gamma_1 -t q^{-2})\xi - \gamma_2 \eta]
    \over P^2 - |Q|^2},\\
y(t; \zeta, \zetab)  &=& { q^2[(1 + \gamma_1 -t )\eta - \gamma_2 \xi]
    \over P^2 - |Q|^2}.
\end{eqnarray}
The higher order configuration invariants are the sums over
the images of products of the signed magnifications with the position
coordinates. The $m$th moment with respect to $x$ and $n$th moment
with respect to $y$ is given by the contour integral
\begin{equation}
\label{eq:highertheoremre}
\sum_{\images} \mu_i p_i x_i^m y_i^n = {q^2 B \over 2
\pi {\rm i}} \oint_\C {dt \over t} { P^2 - |Q|^2 \over K(t; \zeta,
\zetab)} [x(t; \zeta, \zetab)]^m [y(t; \zeta, \zetab)]^n.
\end{equation}
Witt \& Mao (2000) introduced some higher order moments of $x$ and $y$
for the cases when the integers $m$ and $n$ are positive and they
showed that the results for the lensing potentials
(\ref{eq:lensing_pot}) were very simple for the point mass ($\beta
=0$) and isothermal ($\beta =1$) cases.

It is also possible to derive moments with respect to the complex image
coordinates.  The $m$th moment with respect to $z$ and the $n$th
moment with respect to $\zb$ is given by the contour integral
\begin{equation}
\label{eq:highertheoremco}
\sum_{\images} \mu_i p_i z_i^m \zb_i^n = {q^2 B \over 2 \pi {\rm i}}
\oint_\C {dt \over t} { {[P\zeta - Q \zetab ]^m [P\zetab -\Qb
\zeta]^n} \over {K(t; \zeta, \zetab)[P^2 - |Q|^2]^{m+n-1}} }
\end{equation}
For positive moments, the results are of course just a recasting of
(\ref{eq:highertheoremre}), but for reciprocal moments, the results
are different.  In both cases~(\ref{eq:highertheoremre})
and~(\ref{eq:highertheoremco}), the contour $\C$ is chosen to enclose
the poles corresponding to the visible images, but to exclude other
poles in the complex $t$-plane such as those at the zeros of $P^2 -
|Q|^2$ when $m+n>1$.


\section{Invariants for integer $B$}

To begin with, let us study elliptic power-law potentials for which
the exponent $B=2/(2-\beta)$ is an integer. This may strike the
reader as narrow in its scope, but it is not really so. As we will
see in \S 4, results which are exactly true for these cases are also
good approximations for the four bright images of all the power-law
galaxies if the source is sufficiently close to the galactic center.
The simplest cases are $B=1$, or $\beta = 0$, when the potential
approximates that of a point mass plus shear, and $B=2$, or $\beta
=1$, when the potential is isothermal-like and represents a galaxy or
cluster with a flat rotation curve. In both these cases, the imaging
equation (\ref{eq:defnk}) is a quartic and has four roots. When $B=3$
or $\beta = 4/3$, so that the convergence or projected mass density
falls off like $r^{-2/3}$, the imaging equation (\ref{eq:defnk}) is a
quintic and has five roots and there is a weak fifth image. The cases
$B=1,2$ and $3$ are special because there are then no spurious roots
of the imaging equation when the source lies within the central
caustic. There are spurious roots of the imaging equation for $B>3$,
which contribute to the invariants which we calculate. However their
contributions are small and many of them cancel, with the consequence
that the results remain useful even when there are spurious roots.
Cases of integer $B$ avoid the need for a branch cut to define the
$t^B$ term in the imaging equation (\ref{eq:defnk}) as a
single-valued function in the complex plane.

\subsection{The Magnification Invariant}

As the images are given by the zeros of the imaging equation, we have
\begin{equation}
\label{eq:theorem_Bsimple}
\sum_{\images} \mu_i p_i = {q^2 B\over 2 \pi {\rm i}} \oint_\C {dt\over t}
{ P^2 - |Q|^2 \over K(t; \zeta, \zetab)}. 
\end{equation} 
The integrand has simple poles at the zeros of $K(t; \zeta, \zetab)$,
as well as at the origin. The contour $\C$ is traversed in an
anti-clockwise direction and encloses the poles corresponding to
images.

Suppose the integrand is taken over a large circle at infinity
$\C_\infty$. As $t \rightarrow \infty$, the integrand falls off at
least as fast as $O(t^{-3})$, so we have;
\begin{equation}
\oint_{\C_\infty} {dt\over t} { P^2 - |Q|^2 \over K(t; \zeta,
\zetab)} =0.
\end{equation}
Distorting the contour so it encloses the zeros of $K(t; \zeta,
\zetab)$ and the pole at the origin, we have 
\begin{equation}
\label{eq:before}
\sum_{\images} \mu_i p_i + {q^2 B \over 2 \pi {\rm i}} \oint_{\C_\epsilon}
{dt\over t} { P^2 - |Q|^2 \over K(t; \zeta, \zetab)} =0,
\end{equation}
where $\C_\epsilon$ is a small circle of radius $\epsilon$ enclosing
the origin in the anticlockwise direction. As $t \rightarrow 0$, then
$K(t; \zeta, \zetab) \rightarrow -(P_0^2 - |Q|^2)^2,$ where
$P_0=P(t=0)={1 \over 2}[1+q^2+\gamma_1(1-q^2)]$ so that 
\begin{equation}
\label{eq:gsimple}
\sum_{\images} \mu_i p_i = {Bq^2 \over P_0^2 - |Q|^2 } = {B \over 1-
\gamma_1^2 - \gamma_2^2}.
\end{equation}
This remarkably simple result was first obtained by Witt \& Mao
(2000), who deduced it for the $B=1$ and $2$ cases and used
interpolation to extend it as an approximation to all $B$. In fact,
the result (\ref{eq:gsimple}) is exact for $B=3$ as well, provided
the contribution from the fifth de-magnified image is included.

\subsection{The Configuration Invariants: First and Second Moments}

The first moments can be obtained in a similar way. When $m=1, n=0$
the integrand (\ref{eq:highertheoremco}) still has poles only at the
zeros of the imaging equation and the origin. Evaluating the residue
at the origin, we find
\begin{equation}
\sum_{\images} \mu_i p_i z_i = {B[P_0\zeta - Q\zetab] \over q^2(1-
\gamma_1^2 - \gamma_2^2)^2}
\end{equation}
and hence
\begin{equation}
\label{eq:resultone}
\sum_{\images} \mu_i p_i x_i = {B( (1\!-\!\gamma_1)\xi\!-\!
\gamma_2 \eta) \over (1\!-\!\gamma_1^2\!-\!\gamma_2^2)^2},
\qquad\qquad
\sum_{\images} \mu_i p_i y_i = {B(
(1\!+\!\gamma_1)\eta\!-\!\gamma_2 \xi) \over
(1\!-\!\gamma_1^2\!-\!\gamma_2^2)^2}. 
\end{equation} 
Witt \& Mao (2000) give these results in the case of on-axis shear
only ($\gamma_2 =0$), while Dalal \& Rabin (2000) give them for the
case $B=2$ only. Algorithms using the real analysis tend to become
cumbersome with off-axis shear. It is part of the power of the
contour integral method that results can be obtained easily for
general shear. When there is no off-axis shear, the $x$ (or $y$)
moments of the signed magnifications of the images are related only
to the $\xi$ (or $\eta$) offset of the lens.

The second moments are given by $m+n=2$. The integrand
(\ref{eq:highertheoremco}) now has additional simple poles at the two
real points $t_1$ and $t_2$, where
\begin{equation}
\label{eq:tonetwodef}
t_1=P_0+|Q|, \qquad t_2=P_0-|Q|,
\end{equation}
where $P^2 - |Q|^2=0$. Provided that $\gamma_1^2+\gamma_2^2<1$, then
$P_0>0$ and $P^2_0-|Q|^2=q^2(1-\gamma^2_1-\gamma^2_2)>0$, and both
$t_1$ and $t_2$ are positive with $t_1>t_2$ and $t_2<1$. The residues
at these locations give rise to constant terms independent of the
source position in the configuration invariants. To show this, let us
consider the evaluation of one of the second moments ($m=1, n=1$) in
a little detail. At large radii, the integrand falls off at least as
fast as $O(t^{-5})$. If evaluated about a circle at infinity
$\C_\infty$, the contour integral (\ref{eq:highertheoremco}) vanishes.
Distorting the contour so it encloses the zeros of $K(t; \zeta,
\zetab)$ and the poles at $t=0, t_1$ and $t_2$, we have 
\begin{equation}
\sum_{\images} \mu_i p_i z_i \zb_i + {q^2 B\over 2 \pi {\rm i}}
\oint_{\C_\epsilon + \C_{t_1} + \C_{t_2}} { dt \over t} {[P\zeta
-Q\zetab][P\zetab - \Qb\zeta] \over [P^2 - |Q|^2] K(t; \zeta,
\zetab)} =0. \end{equation}
Here, $\C_\epsilon$, $\C_{t_1}$ and $\C_{t_2}$ are small circles of
radius $\epsilon$ enclosing the simple poles at $t=0, t_1$ and $t_2$
respectively. By evaluating the residues, we find; 
\begin{equation}
\label{eq:resulttwoz}
\sum_{\images}\mu_ip_iz_i\bar{z}_i
={q^2B\over 2|Q|}
\left\{
{1\over t^{B+1}_2}
-{1\over t^{B+1}_1}
\right\}
+{B(P_0\zeta-Q\zetab)(P_0\zetab-\bar{Q}\zeta) \over
q^4(1-\gamma^2_1-\gamma^2_2)^3}
\end{equation}
After evaluating the moment of $z_i^2$ in a similar manner, we obtain
the real second moments:
\begin{eqnarray}
\label{eq:resulttwoa}
\sum_{\images}\mu_ip_ix^2_i
&=&{q^2B\over 4|Q|^2}
\left\{
{|Q|-\Re(Q)\over t^{B+1}_2}
-{|Q|+\Re(Q)\over t^{B+1}_1}
\right\}
+ {B( (1\!-\!\gamma_1)\xi\! -\! \gamma_2 \eta)^2 \over (1\!-\!
\gamma_1^2\! -\!\gamma_2^2)^3},\\
\label{eq:resulttwob}
\sum_{\images}\mu_ip_iy^2_i
&=&{q^4B\over 4|Q|^2}
\left\{
{|Q|+\Re(Q)\over t^{B+1}_2}
-{|Q|-\Re(Q)\over t^{B+1}_1}
\right\}
+ {B( (1\!+\!\gamma_1)\eta\!-\!\gamma_2
\xi)^2 \over (1\!-\! \gamma_1^2\! - \!\gamma_2^2)^3},\\
\label{eq:resulttwoc}
\sum_{\images}\mu_ip_ix_iy_i
&=&{-\gamma_2q^4B\over 4|Q|}
\left\{
{1\over t^{B+1}_2}
+{1\over t^{B+1}_1}
\right\}
+{B((1\!+\!\gamma_1)\eta\! -\!
\gamma_2\xi) ((1\!-\!\gamma_1)\xi\! -\! \gamma_2\eta) \over
(1\!-\!\gamma_1^2\!-\!\gamma_2^2)^3}.
\end{eqnarray}
These results simplify considerably when the shear acts only on axis
$(\gamma_2=0)$, in which case $|Q|=\Re(Q)$, $t_1=1+\gamma_1$, and
$t_2=q^2(1-\gamma_1)$. The $x^2$ (or $xy$ or $y^2$) moments of the
signed magnifications of the images are related only to the $\xi^2$
(or $\xi\eta$ or $\eta^2$) offset of the lens. The results
(\ref{eq:resulttwoa}) - (\ref{eq:resulttwoc}) are given by Witt \& Mao
(2000) for the case of vanishing off-axis shear. For completeness, the
third moments are derived in Appendix B.

\subsection{The Configuration Invariants: Reciprocal Moments}

The reciprocal moments can also be found. The first order reciprocal
moments are given by $m=-1, n=0$ or $m=0, n= -1$. In addition to the
poles at the images and the origin, the integrand also has simple
poles at either $t_3$ or $t_4$ given by
\begin{equation}
t_3 = q^2 \left(1\! - \!\gamma_1\! -\! {\gamma_2\eta \over
\xi}\right), \qquad t_4 = 1\!+\! \gamma_1\! -\! {\gamma_2 \xi \over
\eta}.
\end{equation}
The real first order reciprocal moments are 
\begin{equation}
\label{eq:resultm1}
\sum_{\images} {\mu_i p_i\over x_i} = C_1 + {B \over
(1\!-\!\gamma_1)\xi
\!-\!\gamma_2 \eta}, \qquad\qquad
\sum_{\images} {\mu_i p_i\over y_i} = C_2 + {B \over
(1\!+\!\gamma_1)\eta
\!-\!\gamma_2 \xi}.
\end{equation}
The constants $C_1$ and $C_2$ are
\begin{eqnarray}
C_1 & = & {q^2B \gamma_2^2 \over \xi t_4[\xi^2t_4^B-\gamma_2^2]}= {B
\gamma_2^2 \over [(1\!-\!\gamma_1)\xi\!-\!\gamma_2 \eta ][-\gamma_2^2
+ \xi^{2-B}q^{2B}
((1\!-\!\gamma_1)\xi\!-\!\gamma_2 \eta)^B]},
\\ C_2 & = & {B \gamma_2^2 \over 
\eta t_5[\eta^2t_5^B-\gamma_2^2]}= {B \gamma_2^2
\over [(1\!+\!\gamma_1)\eta\!-\!\gamma_2 \xi ][-\gamma_2^2 + \eta^{2-B}
((1\!+\!\gamma_1)\eta\!-\!\gamma_2 \xi)^B]}. 
\end{eqnarray}
Sending $\xi \rightarrow \eta$, $\eta \rightarrow \xi$, $\gamma_1
\rightarrow - \gamma_1$, $q \rightarrow 1$, then we see that
$C_1 \rightarrow C_2$.  This follows as $C_1$ is the residue of the
integrand at $t_3$, and $C_2$ at $t_4$. Note that $C_1$ and $C_2$ both
vanish when the shear is on-axis only ($\gamma_2 =0$). In this
instance, it is again true that the reciprocal $x$ (or $y$) moments of
the signed magnifications depend only on the reciprocal $\xi$ (or
$\eta$) offset of the source.

The complex reciprocal moments have poles of order $-m$ or $-n$ at
$t=t_5=P_0-Q\zetab/\zeta$ and/or its conjugate
$t=\bar{t}_5=P_0-\bar{Q}\zeta/\zetab$. For $m=-1$ and $n=0$, the
residues at $t=0$ and $t=t_5$ exactly cancel so that
\begin{equation}
\label{eq:zrecipone}
\sum_{\images}
{\mu_ip_i\over z_i} =0.
\end{equation}
This implies the vanishing of the following two real moments 
\begin{equation}
\label{eq:zrecipbits}
\sum_{\images}
{\mu_ip_ix_i\over x^2_i+{y^2_i q^{-2}}} =\sum_{\images}
{\mu_ip_iy_i\over x^2_i+{y^2_i q^{-2}}} =0. \end{equation}
A complication that arises with sufficiently negative $m+n$ is that
there is an additional contribution from the integration over the
large circle at infinity $C_{\infty}$ to be included. This happens
when $-(m+n)$ equals or exceeds the larger of $B$ and $2$, and hence
with second order and higher reciprocal moments. The second order
reciprocal moments in complex form have the compact expressions
\begin{equation}
\sum_{\images}{\mu_ip_i\over z_i\bar{z}_i}
=\sum_{\images}{\mu_ip_i\over (x^2_i+y^2_iq^{-2})^2}
=\left\{
\begin{array}
	 {l@{\quad}l}
	 0,& B<2,\\
	 \noalign{\vskip7pt}
	 \displaystyle{2q^2\over|\zeta|^2-1},& B=2,\\
	 \noalign{\vskip7pt}
	 \displaystyle{Bq^2\over|\zeta|^2},&B>2,
\end{array}
\right.
\end{equation}
and
\begin{equation}
\sum_{\images}{\mu_ip_i\over z^2_i}
=\sum_{\images}{\mu_ip_i\over (x^2_i+y^2_iq^{-2})^2}
\left[x^2_i-{y^2_i\over q^2}-{2{\rm i}x_iy_i\over q}\right]
=\left\{
\begin{array}
	 {l@{\quad}l}
	 \displaystyle-{Bq^2\bar{Q}t^{B-1}_5\over|Q|^2},& B<2,\\
	 \noalign{\vskip7pt}
	 \displaystyle 2q^2\left[{\bar{\zeta}^2\over|\zeta|^2-1}
	    -{\bar{Q}t_5\over|Q|^2}\right],& B=2,\\
	 \noalign{\vskip7pt}
	 \displaystyle Bq^2\left[{1\over\zeta^2}
	    -{\bar{Q}t^{B-1}_5\over|Q|^2}\right],&B>2.
\end{array}
\right.
\end{equation}
There are also moments for which one of $m$ and $n$ is negative, and
the other positive. The simplest is that for $m=1$ and $n=-1$, which
gives
\begin{equation}
\sum_{\images}{\mu_ip_iz_i\over\bar{z}_i}
=-{BP_0 \over \Qb(1-\gamma^2_1-\gamma^2_2)},
\end{equation}
or, in real form,
\begin{eqnarray}
\sum_{\images}{\mu_ip_i(x^2_i-y^2_iq^{-2})
    \over x^2_i+y^2_iq^{-2}}
&=&-{BP_0\Re(Q)\over(1-\gamma^2_1-\gamma^2_2)|Q|^2},\nonumber\\
\sum_{\images}{\mu_ip_ix_iy_i\over x^2_i+y^2_iq^{-2}}
&=&-{\gamma_2q^2BP_0\over 2(1-\gamma^2_1-\gamma^2_2)|Q|^2}.
\end{eqnarray}

\begin{figure*}
\plotone{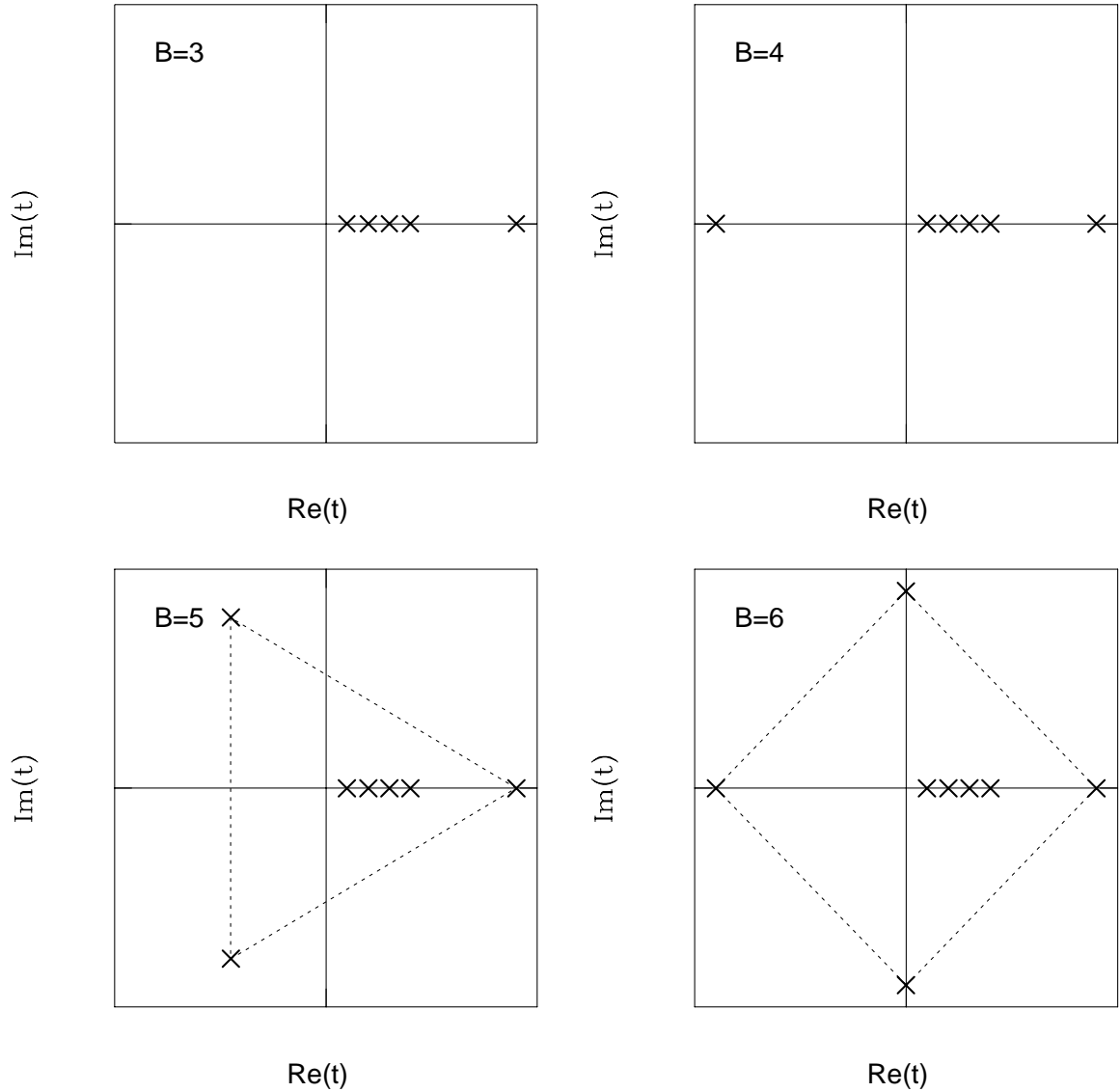}
\caption{The locations of the roots of the imaging equation in the
complex $t$-plane for the cases $B= 3, 4, 5$ and $6$. New spurious
roots appear on the negative real axis whenever $B$ is an even
integer. As $B$ increases, the spurious root bifurcates and the pair
move off into the complex plane with arguments $\exp \pm 2\pi i/
(B\!-\!2)$. For integer $B$, the roots lie at the vertices of a
regular $B\!-\!2$ sided polygon.}
\label{fig:schematic}
\end{figure*}

\subsection{Spurious Roots}

Now let us confront the problem of the spurious roots of the imaging
equation (\ref{eq:defnk}) that do not correspond to physical images.
When $B \le 2$, there are four roots on the real positive $t$-axis
corresponding to the four images of a quadruple lens. When $B
>2$, there are $B\!+\!2$ roots in total. With $|\zeta|$ small, the
$B\!-\!2$ additional roots appear at large $t$ (or at small $|z|$).
By balancing terms in the imaging equation, we see that these roots
$\ts$ satisfy to leading order
\begin{equation}
\label{eq:spurious}
\ts = {1 \over |\zeta|^{2/ (B\!-\!2)}} \exp \left({ 2n \pi {\rm i} \over
B\!-\!2}\right) + O(1),
\end{equation}
where $n$ is an integer so that there are $B\!-\!2$ distinct roots of
this form. Only one of them ($n=0$) lies on the real positive $t$-axis
and corresponds to a physical image. In fact, it is the strongly
de-magnified fifth image that occurs so close to the centre of the
lens that it is often hard to detect. The other $B\!-\!3$ roots
($0<n\leq B\!-\!3$) are spurious and do not provide physical
images. Figure~\ref{fig:schematic} illustrates the behaviour of the
roots in the complex $t$-plane as $B$ increases. The first panel shows
the case $B=3$, with the five roots corresponding to the five images
along the real positive $t$-axis. The second panel shows the case
$B=4$, when a spurious root appears at large negative real $t$.  When
$4 < B <6$, this negative real root splits into a pair with arguments
$\exp \pm 2\pi {\rm i}/ (B\!-\!2)$. In the third panel, at $B=5$, we
have three roots equally spaced around a circle of large radius in the
complex $t$-plane. In the fourth panel, the cycle of events begins
again with a new spurious root appearing on the negative real axis
when $B=6$.

Some small part of the invariants that we calculated earlier in this
section are due to spurious roots. Their contributions can now be
calculated and subtracted. We shall do this for the sum of the signal
magnitudes. In fact, we shall subtract also the contribution from the
weak fifth image to obtain an approximation for the sum of the signed
magnitude of the four bright images. The contribution to the integral
(\ref{eq:theorem_Bsimple}) from the residue at each root
(\ref{eq:spurious}) is
\begin{equation}
\Residue (t = \ts) = {Bq^2 \over (B\!-\!2)} |\zeta|^{4/(B\!-\!2)}
\exp \left( -{4 n \pi {\rm i}\over B\!-\!2} \right) +
O(|\zeta|^{6/(B\!-\!2)}). 
\end{equation} 
For odd values of $B$, it is convenient to set $B = 3\!+\!2N$. The
spurious roots lie at angles $\exp (\pm 2\pi {\rm i} n/(2N\!+\!1))$ and so 
\begin{eqnarray}
\label{eq:oddb}
\sum_{4\,\images} \mu_i p_i &=& {B\over(1-\gamma^2_1-\gamma^2_2)} - 
{B q^2 \over (B\!-\!2)} |\zeta|^{4/(B\!-\!2)}\nonumber\\
&&\times\left[1 + \sum_{n=1}^N \exp\left( {4n\pi {\rm i} \over
2N\!+\!1}\right)+\exp\left(-{4n \pi {\rm i} \over 2N\!+\!1}\right) \right]
+O(|\zeta|^{6/(B\!-\!2)}).
\end{eqnarray}
For even integers $B = 2\!+\!2N$, the spurious roots lie at angles
$\exp (\pm n \pi i /N)$ and so
\begin{eqnarray}
\label{eq:evenb}
\sum_{4\,\images} \mu_i p_i &=& {B\over(1-\gamma^2_1-\gamma^2_2)} - {
B q^2 \over (B\!-\!2)}|\zeta|^{4/(B\!-\!2)}\nonumber\\ &&\times\left[2
+ \sum_{n=1}^{N\!-\!1} \exp\left( {4 n\pi {\rm i}\over N}\right)+
\exp\left(-{4n \pi {\rm i}\over N}\right)
\right]+O(|\zeta|^{6/(B\!-\!2)}).
\end{eqnarray}
Except for the cases $B=3$ and $B=4$, the terms in square brackets on
the right-hand side of (\ref{eq:oddb}) and (\ref{eq:evenb}) sum to
zero, and so there is no contribution to the magnification invariant
at this order. This is not the case for $B=3$, when the sum in
(\ref{eq:oddb}) is empty, and we obtain: 
\begin{equation}
\label{eq:MBthree}
\sum_{4\,\images} \mu_i p_i = {3\over(1-\gamma^2_1-\gamma^2_2)} - 3
q^2 |\zeta|^4 + O(|\zeta|^6),
\end{equation}
or for $B=4$ when the sum in (\ref{eq:evenb}) is empty, and we obtain
\begin{equation}
\label{eq:MBfour}
\sum_{4\,\images} \mu_i p_i = {4\over(1-\gamma^2_1-\gamma^2_2)} - 4
q^2 |\zeta|^2 + O(|\zeta|^4).
\end{equation}
In both cases, the corrections match those given by the more general
theory which we develop in the next section.

\section{Invariants for all $B$}

The general case in which $B$ is no longer an integer requires that a
branch cut be introduced to make single-valued the $t^B$ term in the
imaging equation (11). This apparent complication is a blessing in
disguise. It allows us to generate invariants which are valid for a
continuous range of values of $B$. They are infinite series in the
source coordinates, whose leading terms are just those that we derived
in \S3 --- namely, eqns (\ref{eq:gsimple}) - (\ref{eq:resultone}),
(\ref{eq:resulttwoz}) - (\ref{eq:resulttwoc}), and
(\ref{eq:resultm1}). Importantly, they are also analytic functions of
$B$, and hence are useful for all values of $B$ in the physically
relevant range $1\leq B<\infty$ ($0\leq\beta<2$). This section
explains why this is so. We first derive the magnification invariant
for the case of general $B<2$ (\S4.1) and then the configuration
invariants (\S4.2). In \S4.3 we show how these invariants could also
be derived more laboriously by perturbation theory. Readers happy to
take the results on trust can turn directly to the applications (\S
5).

\subsection{A Recalculated Magnification Invariant}

\begin{figure}
\begin{center}
\plotone{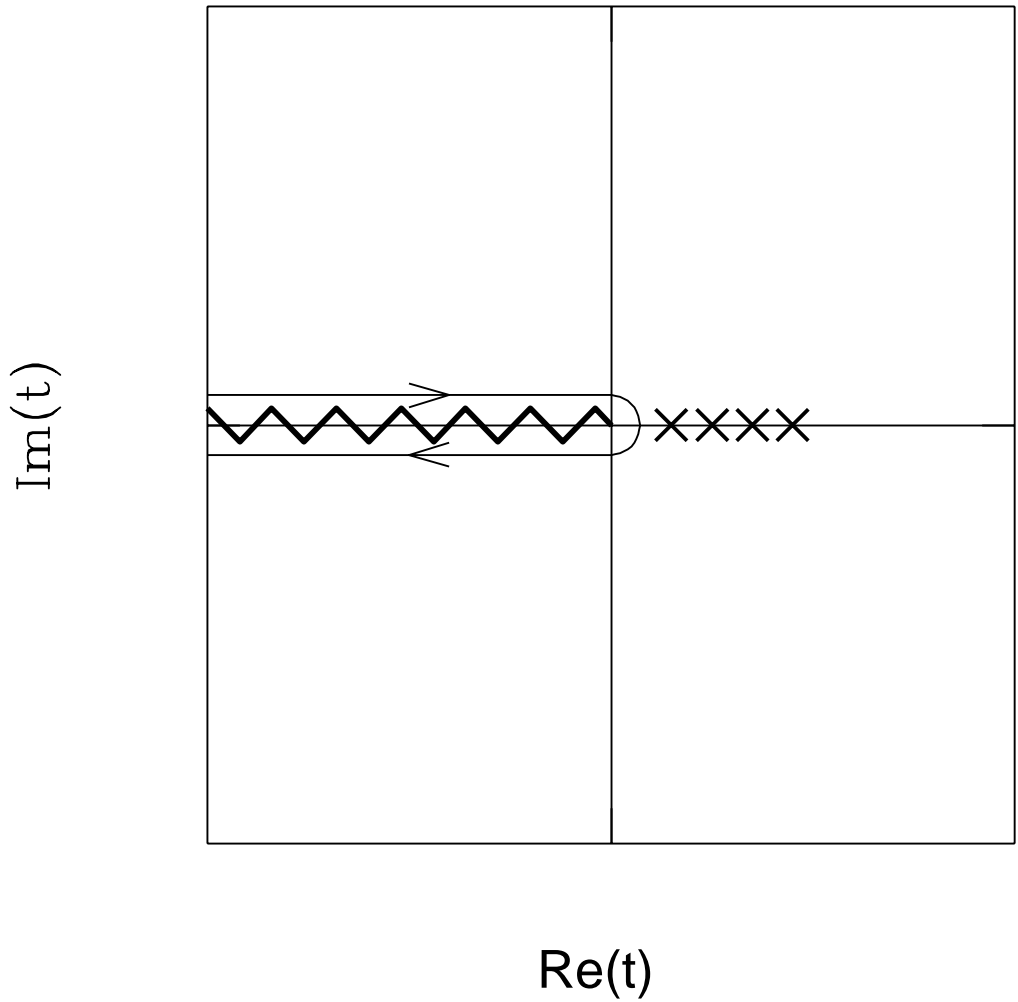}
\end{center}
\caption{The contour for eq. (\ref{eq:cut}) is shown. The $t$-plane
is cut along the negative real axis to make $t^B$ single-valued. The
contour is completed by a large circle at infinity, along which the
integrand is vanishingly small. }
\label{fig:cut}
\end{figure}

Suppose that $B$ is not an integer and that $1\leq B\leq 2$, so that
there are no spurious roots and only the four bright images. We make
$t^B$ single-valued and real on the positive real $t$-axis with a cut
along the negative real axis. We again use the contour integral
(\ref{eq:theorem_Bsimple}) for the magnification invariant. We
enlarge the contour $C$ as much as possible --- the integrand has no
singularities other than those on the cut --- until it consists of
the infinite loop $\Ccut$ shown in Figure~\ref{fig:cut} together with
a large circle at infinity. The large circle again contributes
nothing, because of the decay of the integrand for large $t$. Hence,
the contour integral becomes
\begin{equation}
\label{eq:cut}
\sum_{\images} \mu_i p_i = {q^2 B \over 2 \pi {\rm i}} \int_{\Ccut}
{dt\over t} { P^2 - |Q|^2 \over K(t; \zeta, \zetab)}. 
\end{equation}
Let us write $K = K_1 + K_2$, where
\begin{equation}
\label{eq:defnkone}
K_1 = -(P^2-|Q|^2)^2=-(t\!-\!t_1)^2(t\!-\!t_2)^2, \qquad\qquad K_2 =
t^B(P\zeta-Q\zetab)(P\zetab-\bar{Q}\zeta). \end{equation}
We introduce the following two linear combinations of the complex
source coordinates:
\begin{equation}
\label{eq:deflm}
	 \lambda={1\over 2}\left(\zeta+{Q\zetab\over|Q|}\right),\qquad
\nu={1\over 2}\left(\zeta-{Q\zetab\over|Q|}\right).\qquad
\end{equation}
They have the important properties that
\begin{equation}
\label{eq:lmprops}
	 \bar{\lambda}\nu+\lambda\bar{\nu}=0,\qquad
|\lambda|^2+|\nu|^2=|\zeta|^2,
\end{equation}
and allow us to write
\begin{equation}
P\zeta-Q\zetab
=\lambda(t_2-t)+\nu(t_1-t),\qquad\qquad (P\zeta-Q\zetab)
(P\zetab-\bar{Q}\zeta)
=|\lambda|^2(t_2-t)^2+|\nu|^2(t_1-t)^2. 
\end{equation}
We expand the denominator of our integrand in an infinite series to
get 
\begin{equation}
\label{eq:series}
\int_{\Ccut} {dt\over t} { P^2 - |Q|^2 \over K(t; \zeta, \zetab)} =
-\sum_{j=0}^{\infty} \int_{\Ccut} dt {P^2 - |Q|^2 \over t K_1}
\left({K_2\over -K_1}\right)^j.
\end{equation}
This is valid for sufficiently small $|\zeta|$ because both $|K_1|$
and $|K_2|$ are bounded below on the cut for finite $t$ and
$|K_2/K_1|\to 0$ as $|t|\to\infty$. The cut is unnecessary for the
first $j=0$ term of the series (\ref{eq:series}). It has just a
simple pole at $t=0$ and so can easily be evaluated. The result is
exactly the same as that of \S3.1 as the residue calculation is
unchanged. In the other terms of the integrand, we set
$t=w\exp(\pm\pi {\rm i})$ on the upper and lower sides of the cut to get
\begin{equation}
\sum_{4\,\images} \mu_i p_i
={B\over 1-\gamma^2_1-\gamma^2_2}
+Bq^2\sum^{\infty}_{j=1}{\sin(Bj\pi)\over\pi}\int^{\infty}_0 dww^{jB-1}
{[|\lambda|^2(w+t_2)^2+|\nu|^2(w+t_1)^2]^j
\over(w+t_1)^{2j+1}(w+t_2)^{2j+1}}.
\end{equation}
Binomial expansion of the integrand gives us integrals which are all
of the form
\begin{equation}
\label{eq:eyetwoint}
I_{j,\ell}(t_1,t_2)
={\sin Bj\pi\over\pi}\int^{\infty}_0
{w^{jB-1}dw\over(w+t_1)^{\ell+1}(w+t_2)^{2j-\ell+1}},\qquad 1\leq
B\leq 2,
\end{equation}
for integers $\ell$ in $[0,2j]$. The integrals can all be expressed
in terms of Gaussian hypergeometric functions ${}_2F_1$, using
formula [3.197.1] of Gradshteyn \& Ryzhik (1965) as %
\begin{equation}
\label{eq:eyetwo}
I_{j,\ell}
={-1\over(2j+1)!}
\prod^{2j+1}_{s=1}(jB-s)
{t^{jB+\ell-2j-1}_2\over t^{\ell+1}_1}
{}_2F_1\left(\ell+1,jB;2j+2;1-{t_2\over t_1}\right). \end{equation}
Appendix C gives closed formulae for some of the low order integrals
$I_{j,\ell}$. Binomial expansion of the numerator of the integrand
gives the following infinite series for the magnification invariant
of the four bright images. We henceforth label this as ${\cal M}(B)$
to emphasise that it is an analytic function of $B$; 
\begin{equation}
\label{eq:fourmag}
{\cal M}(B)
=\sum_{4\,\images}\mu_ip_i
={B\over 1-\gamma^2_1-\gamma^2_2}
+Bq^2\sum^{\infty}_{j=1}\sum^j_{m=0}
{j\choose m}I_{j,2m}|\lambda|^{2m}|\nu|^{2j-2m} \end{equation}
The $j$th term is a homogeneous polynomial of degree $2j$ in the
source coordinates. Writing out the first correction term explicitly,
we have
\begin{equation}
\label{eq:fourmagval}
{\cal M}(B)
={B\over 1-\gamma^2_1-\gamma^2_2}
+Bq^2\left[K^{0\,0}_{2\,0}\xi^2
+K^{0\,0}_{1\,1}\xi\eta
+K^{0\,0}_{0\,2}\eta^2\right]
+O(|\zeta|^4),
\end{equation}
where
\begin{eqnarray}
K^{0\,0}_{2\,0}&=&{1\over 2}[I_{1,2}+I_{1,0}] +{\Re\,Q\over
2|Q|}[I_{1,2}-I_{1,0}],\qquad\qquad
K^{0\,0}_{1\,1}={\gamma_2q^2\over|Q|}[I_{1,2}-I_{1,0}],\nonumber\\
K^{0\,0}_{0\,2}&=&{q^2\over 2}[I_{1,2}+I_{1,0}] -{\Re\,Q\over
2|Q|}[I_{1,2}-I_{1,0}],
\end{eqnarray}
and, using eqn (\ref{eq:hypergeomtwo}), we find, 
\begin{eqnarray}
I_{1,2}\!-\!I_{1,0}&=&{(B\!-\!1) \over 4|Q|^2}
\Bigl[t_1^{B\!-\!2}\!-\!t_2^{B\!-\!2}
\!-\!(B\!-\!2)|Q|(t_1^{B\!-\!3}\!+\!t_2^{B\!-\!3})\Bigr],\nonumber\\
I_{1,2}\!+\!I_{1,0}&=&{(B\!-\!1) \over 4|Q|^2}\Bigl[t_1^{B\!-\!2}
\!+\!t_2^{B\!-\!2}\!-\!(B\!-\!2)|Q|(t_1^{B\!-\!3}\!-\!t_2^{B-3})
\Bigr]\!+\!{1 \over 4|Q|^3}\Bigl[t_2^{B-1}-t_1^{B-1}\Bigr].
\end{eqnarray}
This gives explicit formulae for the first order corrections to the
magnification invariant. Once again, the formulae simplify
considerably if there is no off-axis shear and $Q$ is real.

Although the preceding derivation of the infinite series
(\ref{eq:fourmag}) for the magnification invariant is valid only for
$1\leq B\leq 2$, the series itself is not so limited. As we show in
Appendix D, it converges for arbitrarily large values of $B$ for
sufficiently small $|\zeta|$. In fact, the series is the unique
analytic continuation of the magnification invariant ${\cal M}(B)$,
which is an analytic function of $B$, and hence can be used for
$B>2$. Here are two simple tests that give credence to this
remarkable result, the justification of which is given in Appendix D.
First, for $B=3$, $I_{1,2}=I_{1,0}=0$ and there is no $O(|\zeta|^2)$
term in the series (\ref{eq:fourmag}). At the next order, we find
from equation (\ref{eq:eyetwo}) that $I_{2,4}=I_{2,2}=I_{2,0}=-1$ and
hence 
\begin{equation}
{\cal M}(3)={3\over 1-\gamma^2_1-\gamma^2_2}
-3q^2|\zeta|^4+O(|\zeta|^6),
\end{equation}
in agreement with the result (\ref{eq:MBthree}) of \S3.4. Second, let
us set $B=4$. Then $I_{1,2}=I_{1,0}=-1$, and evaluating the $j=1$
term of (\ref{eq:fourmag}) and using (\ref{eq:lmprops}) gives %
\begin{equation}
{\cal M}(4)={4\over 1-\gamma^2_1-\gamma^2_2}
-4q^2|\zeta|^2+O(|\zeta|^4).
\end{equation}
This matches the approximate result (\ref{eq:MBfour}) we
obtained in \S3.4 after subtracting the weak fifth and one spurious
image.

\subsection{Recalculated Configuration Invariants}

We now obtain infinite series expansions for the configuration
invariants (\ref{eq:highertheoremco}), also summed over the four
bright images, by manipulating that contour integral in a manner
similar to that of the previous section. The integrand now generally
has additional finite singularities at one or more of $t_1$, $t_2$,
$t_5$, and ${\bar t_5}$, and so their residues must be accounted for
when they are crossed in the process of enlarging the contour and
wrapping it around the cut. The result is: 
\begin{eqnarray}
\sum_{4\,\images} \mu_i p_iz^m_i{\bar z}^n_i
&=&{B[P_0\zeta\!-\!Q\zetab]^m [P_0\zetab\!-\!{\bar Q}\zeta]^n \over
q^{2m\!+\!2n}[1\!-\!\gamma^2_1\!-\!\gamma^2_2]^{m\!+\!n\!+\!1}}
\nonumber\\
&\!-\!&q^2B\sum_{\poles}\;{\rm{residues\ of}}\left\{
{[P\zeta\!-\!Q\zetab]^m[P\zetab\!-\!{\bar Q}\zeta]^n \over
tK(t;\zeta,\zetab)[P^2\!-\!|Q|^2]^{m+n-1}}\right\}\!+\!S_{m,n}(B),
\end{eqnarray}
where
\begin{equation}
S_{m,n}(B)
=Bq^2\sum^{\infty}_{j=1}
{\sin(Bj\pi)\over\pi}\int^{\infty}_0{dww^{jB\!-\!1}
[\lambda(w\!+\!t_2)\!+\!\nu(w\!+\!t_1)]^{m\!+\!j}
[{\bar\lambda}(w\!+\!t_2)\!+\!{\bar\nu}(w\!+\!t_1)]^{n\!+\!j}
\over[(w\!+\!t_1)(w\!+\!t_2)]^{2j\!+\!m\!+\!n\!+\!1}}. 
\end{equation}
The binomial expansion of the integrand is less simple than in \S4.1,
as $m\not=n$ and a double summation is needed. The result is the
following homogeneous polynomial of $\lambda$, ${\bar\lambda}$,
$\nu$, and ${\bar\nu}$:
\begin{equation}
\label{eq:Smnseries}
S_{m,n}(B)
\!=\!Bq^2\sum^{\infty}_{j=1}\sum^{2j\!+\!m\!+\!n}_{\ell\!=\!0}
I_{j,\ell,m\!+\!n}
\sum^{k=\min(\ell,m\!+\!j)}_{k=\max(0,\ell\!-\!n\!-\!j)}
{m\!+\!j\choose k}{n\!+\!j\choose \ell\!-\!k}
\lambda^k{\bar\lambda}^{\ell\!-\!k}
\nu^{m\!+\!j\!-\!k}{\bar\nu}^{n\!+\!j\!+\!k\!-\!\ell}. \end{equation}
The $j$th term is now a homogeneous polynomial of degree $2j+m+n$ in
the source coordinates. A more general integral with an extra
subscript (which was zero previously) now appears as a coefficient.
It is
\begin{eqnarray}
\label{eq:eyethree}
I_{j,\ell,N}(t_1,t_2)
&=&{\sin Bj\pi\over\pi}\int^{\infty}_0
{w^{jB\!-\!1}dw\over(w\!+\!t_1)^{\ell\!+\!1}
(w\!+\!t_2)^{2j\!+\!N\!-\!\ell\!+\!1}},\qquad
0\leq B\leq 2 \\
&=&{(-1)^{N\!+\!1}\over(2j\!+\!N\!+\!1)!}\prod^{2j\!+\!N\!+\!1}_{s=1}(jB\!-\!s)
{t^{jB\!+\!\ell\!-\!2j\!-\!N\!-\!1}_2\over t^{\ell\!+\!1}_1}
{}_2F_1\left(\ell\!+\!1,jB;2j\!+\!2\!+\!N;1\!-\!{t_2\over
t_1}\right).\nonumber 
\end{eqnarray}
Although the integral definition of $I_{j,\ell,N}$ is valid only if
$B\leq 2$, the hypergeometric formula is valid for all $B>1$.
Explicit closed form expressions for all the hypergeometric functions
are given in Appendix C. Using these results allows us to extend the
formulae given in \S3.2 and \S3.3 to configuration invariants. The
larger $m+n$ is, the longer they are, so we shall here quote only the
first reciprocal moments, which are
\begin{eqnarray}
\sum_{\images}
{\mu_ip_ix_i\over (x^2_i\!+\!y^2_i q^{-2})}\!&=&\!{Bq^2\over 2}
\Biggl[ \Bigl[ \bigl( 1\!-\!{\Re\,Q\over 2|Q|}\bigr)I_{1,0,-1}
\!+\!\bigl( 1\!+\!{\Re\,Q\over 2|Q|} \bigr) I_{1,1,-1} \Bigr]\xi
\!+\!{\gamma_2q^3\eta \over |Q|}[I_{1,1,-1}\!-\!I_{1,0,-1}] \Biggr],
\nonumber\\
\sum_{\images}
{\mu_ip_iy_i\over (x^2_i\!+\!y^2_iq^{-2})}\!&=&\! {Bq^4\over 2}
\Biggl[ \Bigl[ \bigl( 1\!+\!{\Re\,Q\over 2|Q|}\bigr)I_{1,0,-1}
\!+\!\bigl(1\!-\!{\Re\,Q\over 2|Q|}\bigr)I_{1,1,-1} \Bigr]\eta
\!+\!{\gamma_2 \xi \over
|Q|}[I_{1,1,-1}\!-\!I_{1,0,-1}]\Biggr],\nonumber \end{eqnarray}
where
\begin{eqnarray}
I_{1,0,-1}\!&=&\!{t_1t_2^{B\!-\!2}\over 4|Q|^2}
\left[(B\!-\!2)\left({t_2\over t_1}\right)
\!+\!\left({t_2\over t_1}\right)^{2-B}\!-\!B\!+\!1 \right],\nonumber\\
I_{1,1,-1}\!&=&\!{t_2^{B\!-\!1}\over 4|Q|^2}
\left[1\!+\!\left({t_2\over t_1}\right)^{1\!-\!B}
\left[B\!-\!2\!+\!(B\!-\!1)\left({t_2\over t_1}\right)\right]
\right]. 
\end{eqnarray}
The first order corrections for the first moments are given in
Appendix C.

\subsection{Perturbation Analysis}

Invariants, and corrections to them, may also be derived by using
perturbation expansions to solve the imaging equation. Its four main
roots tend in pairs to $t_1$ and $t_2$ as the source tends to the
center of the lensing galaxy, and the expansions are in integer
powers of $\zeta$ and $\zetab$. One pair of images lies at 
\begin{equation}
\label{eq:toneimage}
	 t=t_1\pm t_1^{B/2}|\lambda|+{1\over 2}Bt_1^{B-1}|\lambda|^2
\pm {1\over 2}t_1^{3B/2}|\lambda|
\Bigl[{B(3B-2)|\lambda|^2 \over t_1^2}+{|\nu|^2 \over 4|Q|^2}\Bigr]
+O(|\zeta|^4),
\end{equation}
and the other at
\begin{equation}
\label{eq:ttwoimage}
t=t_2\pm t_2^{B/2}|\nu|+{1\over 2}Bt_2^{B-1}|\nu|^2
\pm {1\over 2}t_2^{3B/2}|\nu|
\Bigl[{B(3B-2)|\nu|^2 \over t_2^2}+{|\lambda|^2 \over 4|Q|^2}\Bigr]
+O(|\zeta|^4),
\end{equation}
where $\lambda$ and $\nu$ are as defined in equation
(\ref{eq:deflm}). The signed magnifications for the first pair of
roots are, from equation (\ref{eq:defmag}),
\begin{eqnarray}
\label{eq:smagone}
\mu p &=& {-q^2 B\over 4|Q|t_1}
\pm {Bq^2|\lambda|t_1^{(B/2)-1} \over 8|Q|}
\Bigl[{1 \over |Q|}-{B-2 \over t_1}\Bigr] \nonumber \\
&& +{Bq^2t_1^{B-2} \over 8|Q|^2}
\Bigl[(B-1)|\lambda|^2 \bigl[1-{(B-2)|Q| \over t_1}\bigr]
-{t_1|\zeta|^2 \over 2|Q|} \Bigr],
\end{eqnarray}
while those for the second pair can be found from them by applying
the transformations $t_1 \to t_2$, $\lambda \to \nu$, and $|Q| \to
-|Q|$. Summation then gives equation (\ref{eq:fourmagval}) again, as
it should.

Finding configuration invariants by perturbation expansion requires
the additional effort of expanding for the image positions $z$ and
${\bar z}$. Though the individual terms are of some intrinsic
interest, perturbation analysis is an unnecessarily laborious way of
computing invariants now that we have the contour integrals which
compute them to all orders.

\section{Astrophysical Applications}

There is inevitable uncertainty in the distribution of mass in lenses,
and all methods (whether parametrised fitting or non-parametric
modelling) must make strong assumptions as to the mass distribution
before they can make progress. The major application of lensing
invariants is to short-cut the modelling process. In \S 5.1, we
provide simple tests to determine whether a given lens system is well
modelled by an elliptic power-law potential with shear. Next, we use
the lensing invariants to develop an algorithm for estimation of the
model parameters and apply it to fake data in \S 5.2 and real data for
the Einstein Cross (G2237+0305) in \S 5.3.  The work in \S 4 has
demonstrated that the simple equations for the moments we derived
in \S3 -- namely, eqns (\ref{eq:gsimple}) - (\ref{eq:resultone}),
(\ref{eq:resulttwoz}) - (\ref{eq:resulttwoc}), and (\ref{eq:resultm1})
-- are excellent approximations for all $B$.  Here, we shall assume
that these results are exact.

\begin{figure}
\begin{center}
\plotone{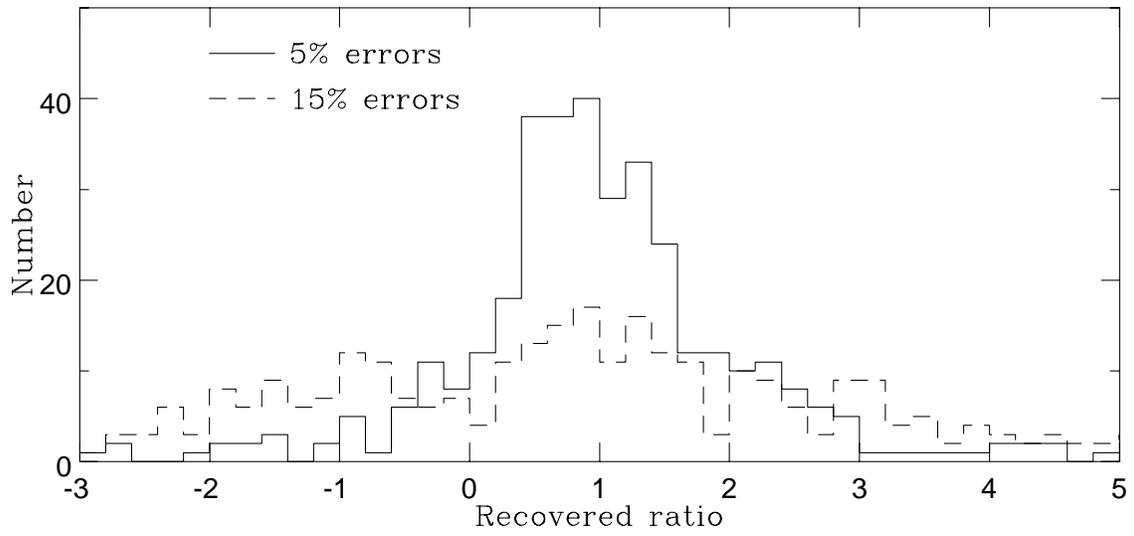}
\end{center}
\caption{The recovery of the test ratios~(\ref{eq:testa}) in the
cases of $5 \%$ and $15 \%$ errors on the photometry. The lens
equation is solved to find the true image positions and fluxes, then
photometry errors are added and the test ratios computed. The
histograms show the distributions obtained with 400 realisations.}
\label{fig:testfig}
\end{figure}
\begin{figure}
\begin{center}
\plotone{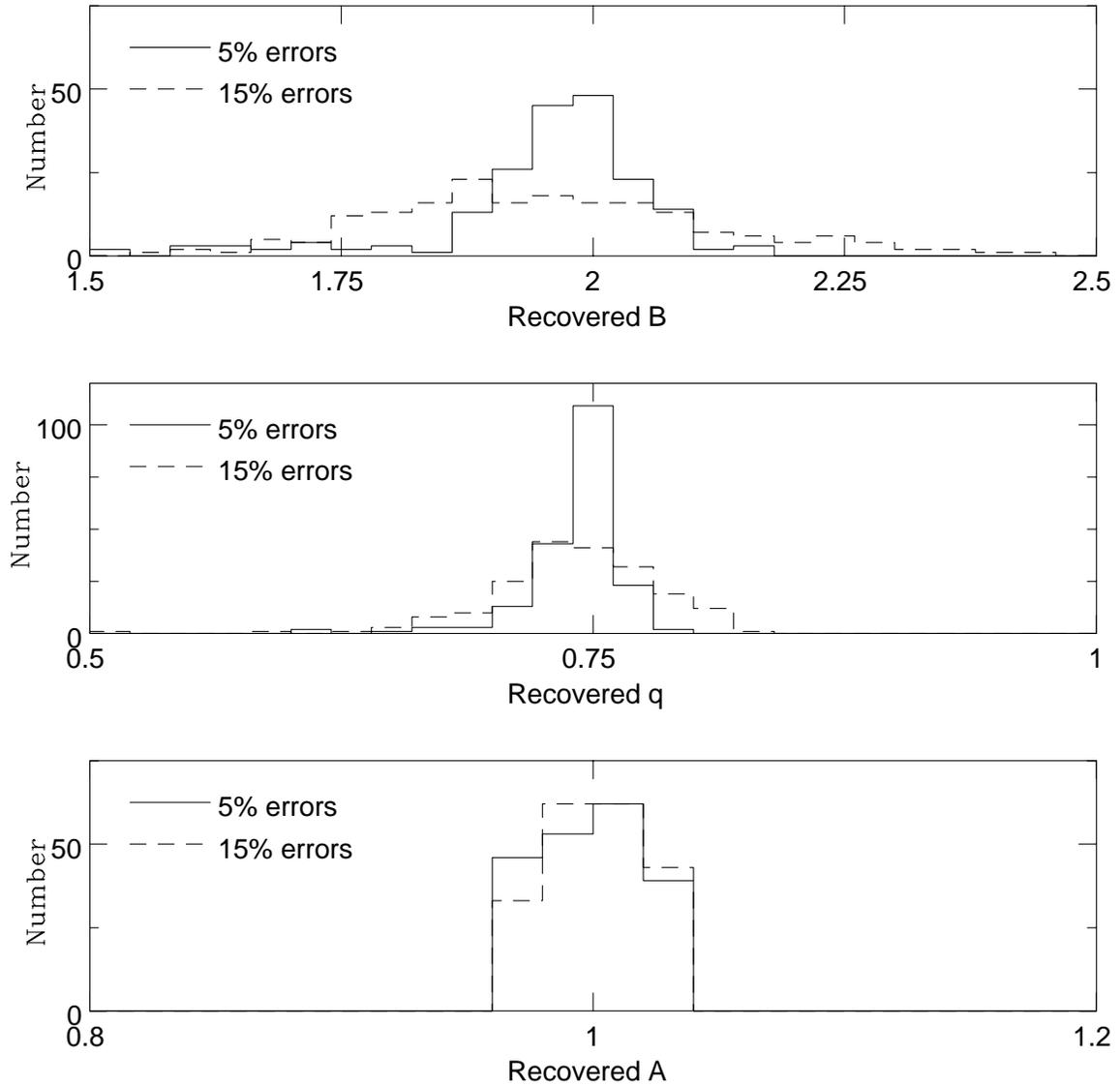}
\end{center}
\caption{The recovery of the axis ratio $q$, the potential
parameters $A$ and $B$ using Monte Carlo simulations. The true values
are $q= 0.75$, $A=1$ and $B=2$. The cases with $5\%$ and $15 \%$
errors on the flux ratios are shown. Notice that $q$ and $A$ are
recovered well, but the larger ($\sim 15 \%$) errors have a
deleterious effect on recovery of $B$. The histograms show the
distributions obtained with 200 realisations.}
\label{fig:monte}
\end{figure}

\subsection{Tests for Elliptic Power-Law Lenses}

From an observational perspective, only the flux ratios of the images
and not their absolute magnifications are measured. The flux ratios
are defined as $f_i = \mu_i/\mu_1$. So, they are normalised to the
flux of the reference image $\mu_1$, which most observers generally
take as the brightest image. For the quadruple lenses, there are three
flux ratios $f_2, f_3$ and $f_4$, with $f_1 = 1$ by definition. So,
the moments of the flux ratios are:
\begin{equation}
\label{eq:fluxmoments}
F_{k,x} = \sum_{i=1}^4 f_ip_ix_i^k,\qquad\qquad F_{k,y} =
\sum_{i=1}^4 f_ip_iy_i^k.
\end{equation}
In other words, $F_0 = F_{0,x} = F_{0,y}$ is just the magnification
invariant normalised by the flux of the reference image $\mu_1$.
Similarly, the higher moments of the flux ratios are given by dividing
the configuration invariants by $\mu_1$.

The source position is just
\begin{equation}
\xi = (1+\gamma_1){F_{1,x}\over F_0} + \gamma_2 {F_{1,y}\over
F_0},\qquad \eta = \gamma_2 {F_{1,x}\over F_0} +
(1-\gamma_1){F_{1,y}\over F_0}. 
\end{equation}
Substituting these into the first moments (\ref{eq:resultone}) and the
reciprocal first moments (\ref{eq:resultm1}), we obtain
\begin{equation}
\label{eq:testa}
{F_{-1,x} F_{1,x} \over F_0^2} = 1,\qquad
{F_{-1,y}F_{1,y}\over F_0^2} = 1. 
\end{equation}
Provided the lensing galaxy can be identified and the position angle
of its major axis can be measured, then all quantities in eqn
({\ref{eq:testa}) are directly available from the data, So, these
equations provide two simple, independent checks that are easy to
apply. If they are satisifed, then the lens may be well-represented by
an elliptic power-law potential with shear. If the major axis of the
lensing galaxy is not known, then one of equations can be used to
estimate the position angle and the second used to check that the
elliptic power-law potential is a reasonable model.

The main difficulty in application of the lensing invariants is the
uncertainty in the observable quantities. The relative positions of
the lensing galaxy and the images may be accurate to better than $1
\%$. In propitious circumstances, the relative fluxes may be good to
perhaps $5 \%$.  Unhappily, microlensing may render these fluxes much
more uncertain, with the error perhaps even as high as $50 \%$ in the
optical (Schechter 2000).  Microlensing is of course valuable, as it
sets constraints on the mass fraction in compact objects in the haloes
of the lensing galaxies (e.g., Witt, Mao \& Schechter 1995). However,
for all researchers modelling the smooth, large-scale distribution of
mass in the lensing galaxy, microlensing is a serious encumbrance. It
is sometimes claimed that microlensing is primarily a difficulty in
the optical wavebands and may be circumvented using radio fluxes.
However, there have been very recent identifications of microlensing
events in the radio (Koopmans \& de Bruyn 2000), and so even the radio
fluxes may be more uncertain than is widely believed. As benchmark
values for our calculations, we take $5 \%$ and $15 \%$ errors.  In
some cases, the radio fluxes may already be accurate to $5 \%$; in
other cases, this accuracy will be achievable in the near future, once
the effects of microlensing are better understood and quantified.  So,
$15 \%$ is a conservative value for the errors in present-day radio
flux measurements.

Figure~\ref{fig:testfig} shows histograms of the distribution of the
ratios~(\ref{eq:testa}) in 400 cases, allowing for $5 \%$ and $15 \%$
errors in the flux measurements. In both cases, the distributions are
peaked around unity, but there is substantial scatter. Even for
datasets with flux errors of $5 \%$, the test should be used in the
form
\begin{equation}
{F_{-1,x} F_{1,x} \over F_0^2} \approx 1 \pm 2,\qquad
{F_{-1,y}F_{1,y}\over F_0^2} \approx 1 \pm 2. 
\end{equation}
for practical application. If the flux errors are as high as $15 \%$,
the tests become much less reliable. If a lens system passes this test
(as for example the Einstein Cross does), then it is worthwhile to
proceed to estimation of the lens parameters.

\begin{table*}
\begin{center}
\begin{tabular}{lccccc} \hline
\null & $-\Delta \alpha$ & $\Delta \delta$ & Radio fluxes & Radio flux \\ 
\null & (in ${}^{\prime\prime}$)
& (in ${}^{\prime\prime}$) & (in $\mu$ Jy) & ratios\\ 
\hline
A & $-0.09$ & $-0.94$ & $65.5$ & $1.00$ \\ 
B & $0.58$ & $0.74$  & $64.2$ & $0.98$ \\ 
C & $-0.72$ & $0.27$ & $26.5$ & $0.40$ \\ 
D & $0.76$ & $-0.42$ & $59.4$ & $0.91$ \\ \hline
\end{tabular}
\end{center}
\caption{Observational data on the Einstein Cross. The positions of
the images are relative to the centre of the lensing galaxy and taken
from Crane et al. (1991). The radio fluxes are provided by Falco, as
reported by Keeton \& Kochanek (1996).}
\end{table*}
\begin{table*}
\begin{center}
\begin{tabular}{cc} \hline
Parameter & Best Fit Value \\ \hline
$A$ & $0.762$  \\ 
$q$ & $0.885$ \\ 
$\beta$ & $0.116$ \\
$\gamma_1$ &$0.117$ \\ 
$\Theta$ & $67.25^\circ$ \\ 
$\xi$ & $-0.1062$ \\
$\eta$ & $-0.0266$ \\ \hline
\end{tabular}
\end{center}
\caption{Estimates of the model parameters for the lensing galaxy of
the Einstein Cross.}
\end{table*}
\begin{table*}
\begin{center}
\begin{tabular}{lccc} \hline
\null & $-\Delta \alpha$ & $\Delta \delta$ & Flux \\ 
\null & (in ${}^{\prime\prime}$)
& (in ${}^{\prime\prime}$) &  ratios\\ \hline 
A & $-0.10$ & $-0.92$ & $1.00$ \\ 
B & $0.58$ & $0.76$ & $0.90$ \\ 
C & $-0.72$ & $0.26$ & $0.40$ \\ 
D & $0.77$ & $-0.40$ & $0.83$ \\ \hline
\end{tabular}
\end{center}
\caption{Model fit to the Einstein Cross using the radio data. This
can be compared with the data given in Table 1.}
\end{table*}

\subsection{Monte Carlo Simulations}

There are three galaxy model parameters, namely $A, q$ and $\beta$. In
the general case, there are two unknown components of shear $\gamma_1$
and $\gamma_2$.  Finally, there are three more unknowns, namely the
source offsets $\xi$ and $\eta$ and the position angle $\Theta$ of the
lensing galaxy.  It is possible to solve formally for the model
parameters, given the magnification invariant and the first, second
and third moments. However, this is not a useful way too proceed as
small errors in the observables amplify into larger errors in the
first, second and third moments. This is exacerbated by the plus and
minus pattern imposed by the parities. The fitting of lens models to
data normally proceeds by the minimisation of the differences between
the observed and true positions (and flux ratios, if desired) using a
downhill simplex routine to search parameter space (e.g., Kochanek
1991). The lensing invariants can be fruitfully exploited to reduce
the dimensionality of the parameter space in which the search
proceeds.

As an example, let us consider a simple case. Suppose that we assume
that the lensing galaxy has negligible external shear. First, it is
straightfoward to find the source offsets
\begin{equation}
\xi = {F_0 \over F_{-1,x}},\qquad \eta = {F_0 \over F_{-1,y}}.
\end{equation}
From the reciprocal second moments, it is then easy to see that the
flattening of the potential is
\begin{equation}
\label{cruxb}
q^2 = {F_{-1,y} \over F_{-1,x}} \sqrt{ {F_{-2,x}F_0 - F_{-1,x}^2
\over F_{-2,y}F_0 - F_{-1,y}^2} }.
\end{equation}
From the second moments, we obtain
\begin{equation}
\label{cruxa}
q^{2B} = -{F_{2,x} F_0 - F_{1,x}^2 \over F_{2,y} F_0 - F_{1,y}^2}.
\end{equation}
from which $B$ can be estimated.  There remain an unknown position
angle $\Theta$ of the lensing galaxy and an unknown potential
normalisation, which can be obtained by minimizing
\begin{equation}
\chi^2 (A, \Theta) = 
\sum_{\images} \Bigl[ \xi - x_i(1 - {A\over (x_i^2 +
 y_i^2q^{-2})^{1 - \beta/2}})\Bigr]^2 + \Bigl[ \eta - y_i(1 - 
{Aq^{-2}\over (x_i^2 +  y_i^2q^{-2})^{1- \beta/2}})\Bigr]^2 
\end{equation}
Figure 3 shows the results of Monte Carlo simulations using fake
data. Taking $A=1$, $B=2$ and $q = 0.75$, the lens equation is first
solved to find the true image positions and fluxes for random source
offsets. A random rotation is performed to take the image positions
from the intinsic coordinate system of the lensing galaxy to the
observer's coordinate system on the plane of the sky. The fluxes are
adusted by random errors of $5\%$ and $15 \%$ to give the simulated
data. This is fed into the algorithm and the random position angle
$\Theta$ and the parameters of the lensing galaxy $A, B$ and $q$ are
recovered. Figure 4 shows histograms of 200 such trials. We see that
if the relative fluxes are accurate to $5 \%$ or even $15 \%$, then
there is good recovery of the model parameters $q$ and $A$. The
parameter controlling the fall-off the potential $B$ seems more
vulnerable to errors, as there are tails to the distribution for
the case of $15 \%$ errors.  Tests also show that essentially no
useful information can be extracted if the fluxes are only accurate to
$50 \%$.

\subsection{The Einstein Cross}

Table~1 presents the available astrometric data on the Einstein Cross
quadruple system from Crane et al. (1991).  The astrometry is provided
in terms of the relative right ascension and declination (in
arcseconds) with respect to the center of the lensing galaxy.  The
radio data on the fluxes of the four images A,B,C and D is given,
rather than the optical data which is known to be strongly effected by
microlensing (e.g., Ostensen et al. 1994). The bulge of the lensing
galaxy has a measured position angle of $68^\circ$ on the sky (Rix et
al. 1992).  If the misalignment of the source is not too great, it is
natural to expect the images of negative parity are located roughly
parallel to the major axis, those of positive parity are roughly
perpendicular to the major axis. Accordingly, we reckon that the
negative parity images are C and D, whilst the positive parity ones
are A and B.

To fit an elliptical power-law model in the presence of external
on-axis shear is straightforward.  We use the relations
\begin{equation}
\label{eq:mina}
\xi = {F_0 (1+ \gamma_1)\over F_{-1,x}},\qquad \eta = {F_0
(1-\gamma_1) \over F_{-1,y}}, \qquad q^2 \Bigl( {1- \gamma_1 \over 1 +
\gamma_1} \Bigr) = {F_{-1,y} \over F_{-1,x}} \sqrt{ {F_{-2,x}F_0 -
F_{-1,x}^2 \over F_{-2,y}F_0 - F_{-1,y}^2} },
\end{equation}
to minimise the $\chi^2$
\begin{equation}
\chi^2 (\gamma_1, A, \beta, \Theta) = 
\sum_{\images} \Bigl[ \xi - x_i(1+ \gamma_1 - {A\over (x_i^2 +
 y_i^2q^{-2})^{1 - \beta/2}})\Bigr]^2 + \Bigl[ \eta - y_i(1- \gamma_1 - 
{Aq^{-2}\over (x_i^2 +  y_i^2q^{-2})^{1- \beta/2}})\Bigr]^2 
\end{equation}
over $\gamma_1$, $A$, $\Theta$. The results for the model parameters
are given in Table 2.  First, let us note that the recovered position
angle agrees well with the observed position angle of the major axis
of the lensing galaxy. Second, only mild on-axis shear is required.
The flattening of the lensing galaxy is significant, but not
unreasonable. Third, the lensing galaxy is not close to isothermal,
and there are no good fits with $\beta =1$.  As all the images are at
almost the same projected distance, the density fall-off, and hence
the parameter $\beta$, is not well-constrained. The total mass
enclosed within the images is more robust and comparable to the values
found by previous investigators (e.g., Rix et al. 1992).  Table 3
shows how well the best fit reproduces the data. The positions and the
relative brightnesses of the four images are reasonably well
reproduced. There are mild discrepancies in the magnifications of
images B and D, which are both somewhat fainter in our model. The
computational advantage of the lensing invariants (\ref{eq:mina}) is
that the $\chi^2$ minimisation can proceed in fewer dimensions, with
obvious gains in accuracy and speed.

Note that Witt \& Mao (2000) give a cautionary example of a model with
a flat rotation curve and with a projected density distribution
stratified on ellipses. Although superficially similar to the elliptic
power-law models, the lensing invariants are quite different. For
example the sum of the signed magnifications is $\sim 2.8$ rather than
2. The magnifications depend on the second derivatives of the lensing
potential and so are more susceptible to the details of the modelling
than the image positions.  Real lenses certainly do exhibit diverse
behaviour. Some -- for example, B1422+231 where the lens is both a
galaxy and a cluster or B1608+656 where the lens is a pair of
interacting galaxies -- are not likely to be well described by any
simple models. Nonetheless, much of the optical depth is in isolated
giant elliptical galaxies or in spirals dominated by dark haloes. It
is reasonable to model such lenses with simple potentials and the
algorithms we have developed have a real r\^ole to play.

\section{Conclusions}

This paper has presented a powerful new method for obtaining lensing
invariants. Such invariants are calculable from the observables and
they remain unchanged as the source of radiation is moved within the
central caustic. The magnification invariant is the sum of the signed
magnifications of the images, whereas the configuration invariants
are sums over powers (either positive or negative) of the image
positions multiplied by the signed magnifications. The idea behind
the method is to use Cauchy's theorem and the residue calculus to
recast sums over the images as contour integrals.

We have illustrated the method by application to a simple family of
galaxy models, the power-law models in which the gravitational
potential is scale-free and stratified on similar concentric
ellipsoids. They may be embedded in an external shear field of
arbitrary orientation. The convergence $\kappa$ of the models falls
off with distance like a power-law with index $-2/B$. The lensing
invariants are exact for the models with $B=1$ (the point mass case),
$B =2$ (an isothermal galaxy with projected potential $\psi \propto
r$) and $B=3$ (a galaxy with $\psi \propto r^{4/3}$). Some of the
lensing invariants for the power-law models were previously
calculated by Witt \& Mao (2000) for the case of on-axis shear. Our
results complement theirs by extending the invariants into the
r\'egime of arbitrary shear, by giving general and more extensive
formulae, as well as by providing some entirely new invariants (the
reciprocal moments).

In their discovery papers, Witt \& Mao (1995, 2000) used direct
elimination methods to find the first examples of lensing invariants.
We believe that our contour integral method offers considerable
advantages over this. All that is needed is the identification of the
poles and the branch points of the lensing equation and the
calculation of the residues using Cauchy's theorem. Our method is
simpler than the multi-dimensional residue calculus proposed by Dalal
\& Rabin (2000). Nevertheless, the idea motivating Dalal \& Rabin's
work -- namely, that sums over images may be replaced by contour
integrals -- is the same as that behind our paper. Only the familiar
single-variable calculus is required in our formulation of the
problem, but it has allowed us to go beyond the polynomial equations
and meromorphic functions studied by Dalal \& Rabin, and to handle
branch points. In turn, this led us to our discovery of full series
expansions for the lensing invariants of the four bright sources,
unsullied by spurious roots of the imaging equation.

We have shown how the lensing invariants may be exploited in modelling
by using Monte Carlo simulations to demonstrate the feasibility of the
recovery of parameters, such as the flattening and the normalisation
of the potential well. Most lens fitting is conventionally done by
$\chi^2$ minimisation. The lensing invariants play an important r\^ole
by reducing the dimensionality of the space in which the global
minimum is sought, with consequent gains in accuracy and speed.  The
power-law models with shear are sensible ones to use when the lens is
a reasonably isolated spiral or elliptical galaxy. To illustrate this,
we have used the lensing invariants to provide a good fit to the
astrometry and photometry of the Einstein Cross (G2237+0305).

Of course, caution is needed in application of the lensing invariants
identified in this paper, as they only apply to a particular family
of galaxy models. The pressing problem is to calculate the lensing
invariants for more general potentials, such as galaxy models with
cores or density distributions stratified on ellipses (e.g. Kassiola
\& Kovner 1993, Barkana 1998). This will lead to a greater
understanding of how the invariants can constrain the mass
distribution for real multiply-lensed systems. We believe that our
contour integral method can be adapted to these cases and are
presently pursuing further investigations.

\bigskip
\noindent
We are indebted to Hans Witt for useful conversations and much
encouragement. NWE thanks the Royal Society for financial support. CH
thanks the sub-department of Theoretical Physics, University of
Oxford for hospitality during a stimulating six-month sabbatical
visit, and the Florida State University for awarding that sabbatical.
His work is supported in part by NSF through grant DMS-9704615.

\bibliographystyle{mnbib}
\bibliography{biblio}

\setcounter{section}{0}
\setcounter{equation}{0}
\begin{appendix}
\section{Images and Caustics}
When the source is sufficiently close to the center of the lens and
$|\zeta|$ small, then there are four images if $B \leq 2$ and five
images if $B>2$. To see this, we split $K$ into two components $K =
K_1 + K_2$, as defined in eqn~(\ref{eq:defnkone}). We recall that,
provided that $\gamma_1^2+\gamma_2^2<1$, both $t_1$ and $t_2$ (defined
in eqn~(\ref{eq:tonetwodef})) are positive with $t_1>t_2$ and
$t_2<1$. As Figure~\ref{fig:chrisone} shows, the graph of $-K_1$, with
its two zero minima at $t=t_1$ and $t=t_2$, has two nearby pairs of
intersections with the positive $K_2$, which is small when $|\zeta|$
is small. There is a fifth image for large positive $t$ when $B>2$
because, however small $|\zeta|$, the asymptotic $t^{B+2}|\zeta|^2$
growth of $K_2$ eventually overtakes the $t^4$ growth of $-K_1$. This
fifth root is given by the $n=0$ case of equation (\ref{eq:spurious}).

The number of images changes when pairs of roots of the imaging
equation merge and become double. Both $K$ and $K_t$ vanish at double
roots, and then 
\begin{equation}
\label{eq:doubleroot}
{1 \over K_2}{\partial K_2 \over \partial t} +{1 \over K_1}{\partial
K_1 \over \partial t} ={B \over t}+{2 \over P-|Q|}+{2 \over P+|Q|}
-{\zeta \over P\zeta-Q\zetab}-{\zetab \over P\zetab-\bar{Q}\zeta}=0.
\end{equation}
Because each denominator is linear in $t$, equation
(\ref{eq:doubleroot}) gives a quartic in $t$, degenerating to a cubic
in the special case of $B=2$. Moreover, the equation is independent of
$|\zeta|$, and depends only on the angular argument $\phi$ of the
complex source position $\zeta=|\zeta|e^{{\rm i}\phi}$. Hence the
source positions which give double images are found by looking for
positive roots of equation (\ref{eq:doubleroot}) along each specific
angular direction. We demonstrate below that, with our restrictions on
the lens parameters, eqn (\ref{eq:doubleroot}) has one real root in
$[t_2,t_1]$ for all $B$, one real root in $[t_1,\infty)$ for $B>2$,
and no others.  Once $t$ is found the imaging equation gives
$|\zeta|^2=[P^2-|Q|^2]^2/[t^B(P-Qe^{-2{\rm i}\phi})(P-\Qb e^{2{\rm
i}\phi})]$. The root in $[t_2,t_1]$ corresponds to the one point on
the tangential caustic in the source direction $\phi$, and the root in
$[t_1,\infty)$ for $B>2$ corresponds to the one point on the radial
caustic in that direction.  These roots must generally be found
numerically, but there are four special directions for which there are
analytical formulas. These are the directions for which $Qe^{-2{\rm
i}\phi}$ is real. When $e^{2{\rm i}\phi}=Q/|Q|$, then $t=t_2$ where
$P=|Q|$ is a triple root of the quartic, and gives a cusp at
$|\zeta|=2|Q|/t_2^{B/2}$ on the tangential caustic. The remaining root
is $t=Bt_1/(B-2)$, which for $B>2$ gives a point
$|\zeta|=(2/B)[(B-2)/Bt_1]^{(B-2)/2}$ on the radial caustic. When
$e^{2{\rm i}\phi}=-Q/|Q|$, then $t=t_1$ where $P=-|Q|$ is a triple
root of the quartic, and $t=Bt_2/(B-2)$ is the fourth root. The former
gives a cusp on the tangential caustic at which
$|\zeta|=2|Q|/t_1^{B/2}$ for $B \leq 2$, and also for $B >2$ if $t_1$
is the smaller of the roots. In the latter case, the fourth root gives
a point $|\zeta|=(2/B)[(B-2)/Bt_2]^{(B-2)/2}$ on the radial
caustic. However, if $Bt_2/(B-2)<t_1$, then the latter point is on the
tangential caustic, while $|\zeta|=2|Q|/t_1^{B/2}$ on the radial
one. Depending on the parameters, either caustic may lie wholly inside
the other, or they may intersect as in the case shown in
Figure~\ref{fig:christwo}.
\begin{figure}
\begin{center}
\plotone{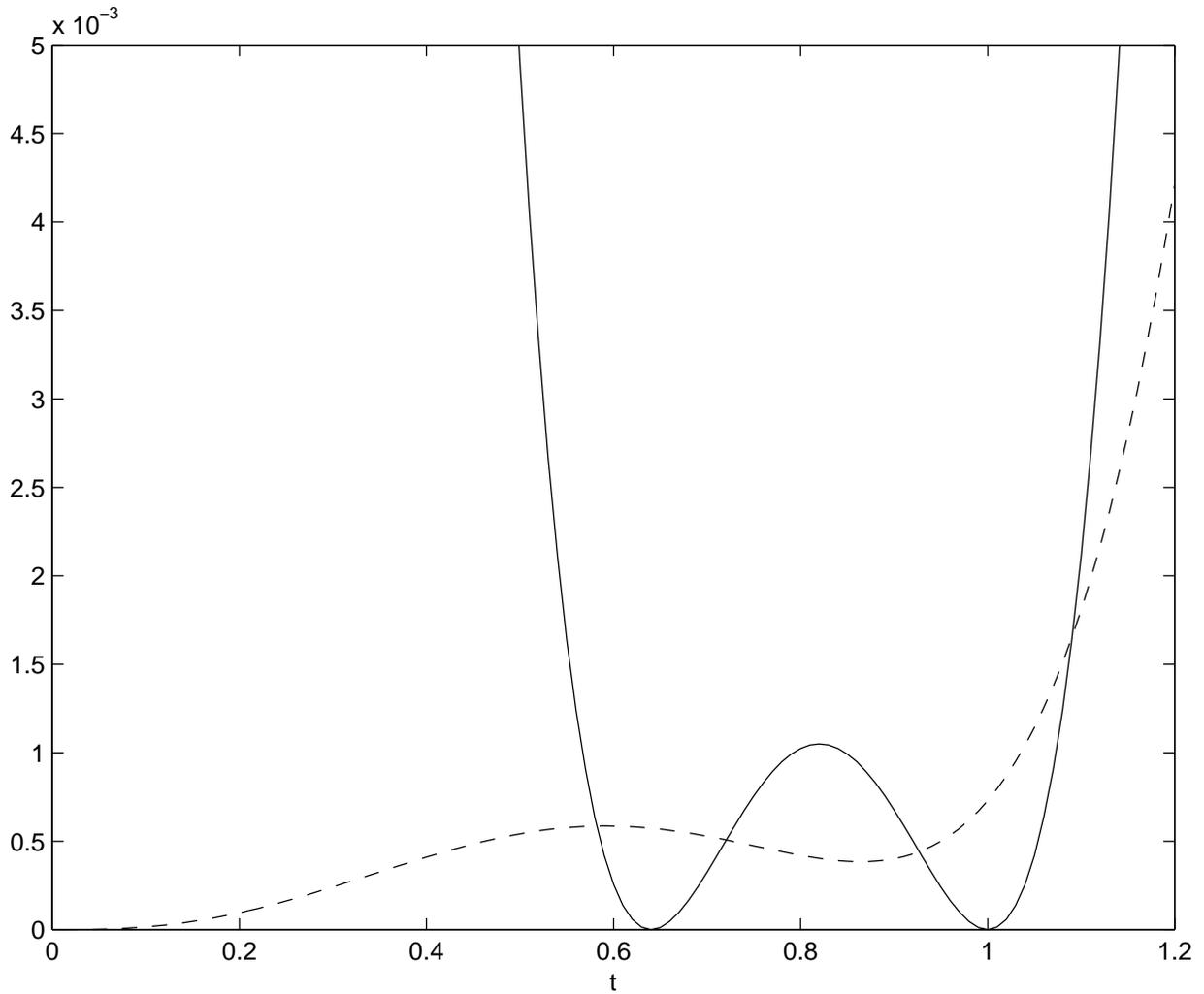}
\end{center}
\caption{Graphs of $K_2(t)$ (dashed curve) and $-K_1(t)$ (full curve)
for the case $B=3$, $q=0.8$, no shear and $\zeta=0.15\exp ({{\rm i}\pi
/3})$. The two curves intersect a fifth time at a much larger value of
$t$. The intersections correspond to the five images.}
\label{fig:chrisone}
\end{figure}
\begin{figure}
\begin{center}
\plotone{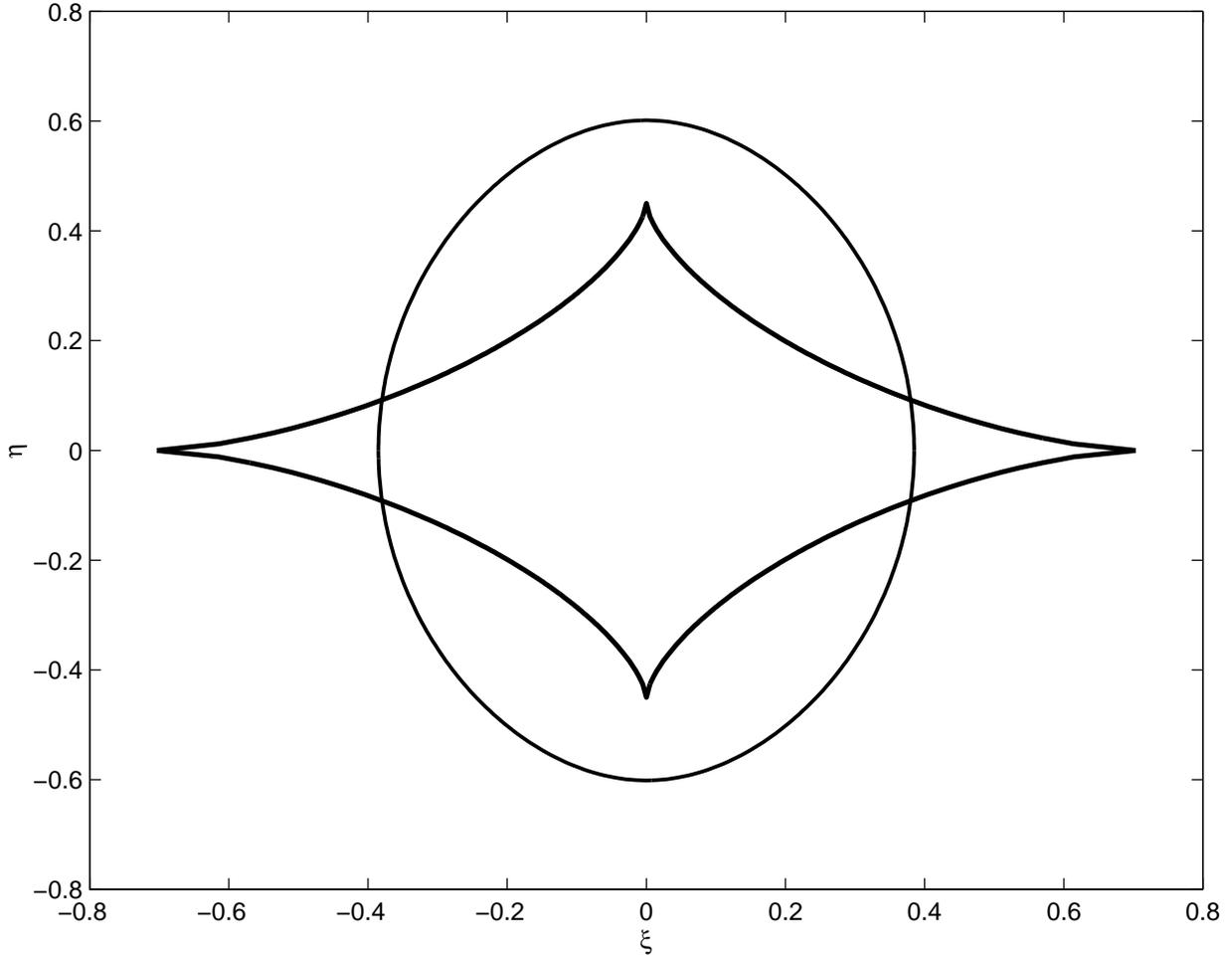}
\end{center}
\caption{The two caustics for the lens of
Figure~\ref{fig:chrisone}. Note that the cusps of the tangential
caustic are ``naked'' and protrude beyond the radial caustic (see
e.g., Evans \& Wilkinson (1998)). }
\label{fig:christwo}
\end{figure}
\begin{figure}
\begin{center}
\plotone{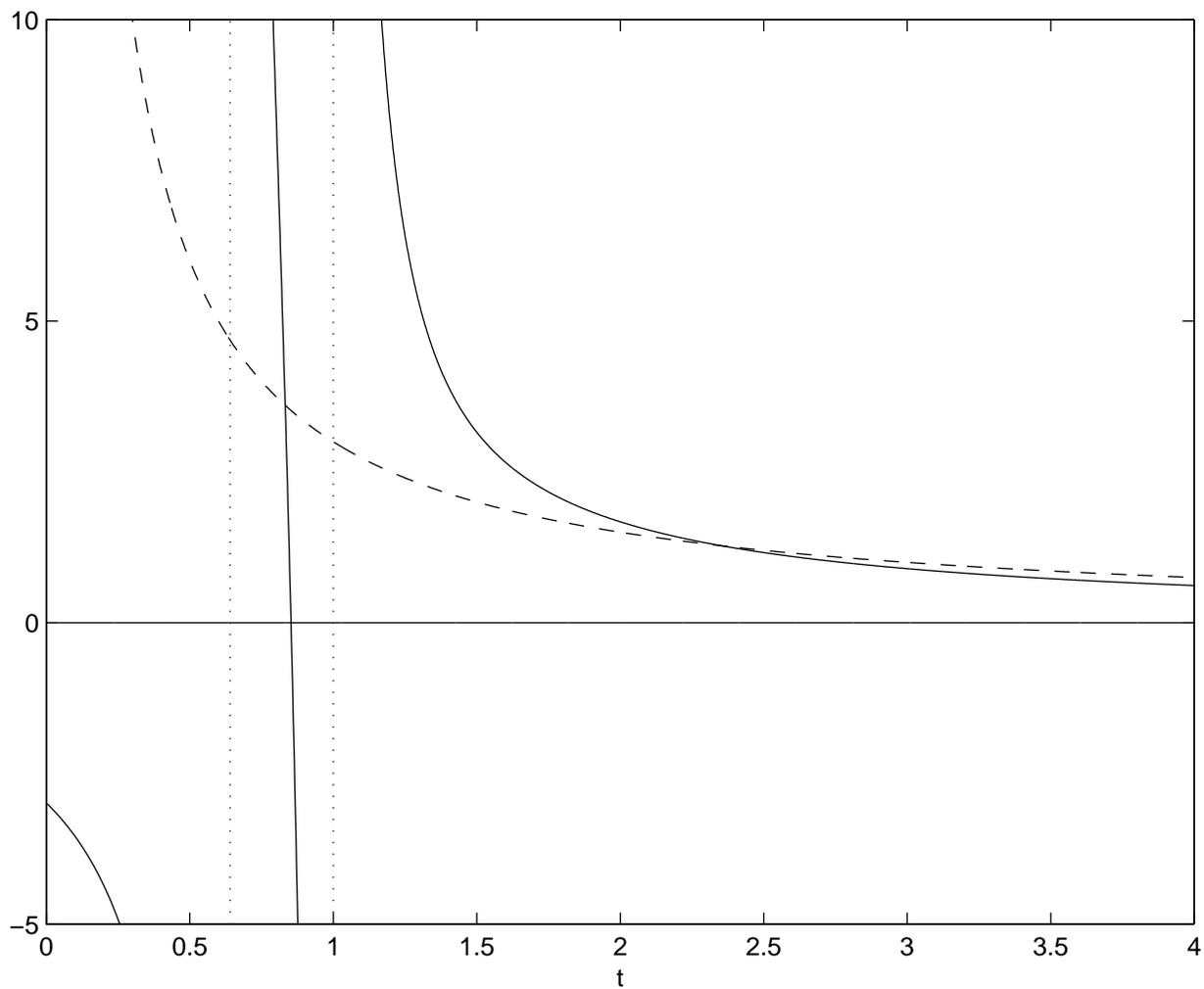}
\end{center}
\caption{
$B/t$ (dashed curve) and the right hand side of equation
(\ref{eq:doubleroottwo}) (full curve) for the case $B=3$, $q=0.8$, and
no shear. The caustics are shown in Figure~\ref{fig:christwo}. The
angle $\theta=\pi/3$.}
\label{fig:christhree}
\end{figure}

In the general case, $Q\zetab/\zeta=|Q|e^{{\rm i}\theta}$ for some
angle $\theta$ which is not an integer multiple of $\pi$. To find the
roots of equation (\ref{eq:doubleroot}) for $-1<\cos\theta<1$, we
rewrite that equation as
\begin{equation}
	 \label{eq:doubleroottwo}
	 {B\over t}={2(P\!-\!|Q|\cos\theta)\over 
          P^2\!+\!|Q|^2\!-\!2P|Q|\cos\theta}
	             +{2\over t\!-\!t_1}+{2\over t\!-\!t_2}
={-2P^3\!+\!6P^2|Q|\cos\theta\!-\!6P|Q|^2\!+\!2|Q|^3\cos\theta
				    \over 
(P^2\!-\!|Q|^2)(P^2\!+\!|Q|^2\!-\!2P|Q|\cos\theta)}.
\end{equation}
Graphs of its two sides as functions of $t$ are shown in
Figure~\ref{fig:christhree}. The right hand side has vertical
asymptotes at $t=t_2$ and $t=t_1$, and a zero in between. This is the
only zero of the numerator because its derivative with respect to $t$,
which is the negative of that with respect to $P$, is
$6[P-|Q|\cos\theta]^2+6|Q|^2\sin^2\theta$ and hence always positive.
The two sides of equation (\ref{eq:doubleroottwo}) have different
signs and hence no roots in $(0,t_2))$. They have at least one root in
$(t_2,t_1)$, and another in $(t_1,\infty)$ for $B>2$ because the right
hand side, which decays as $2/t$ is ultimately the smaller for large
$t$.  To prove that these are the only possibilities, we rewrite
equation (\ref{eq:doubleroottwo}) as a quartic in two different ways,
and apply Descartes' rule of signs (Henrici 1974). We also reduce the
number of parameters to three by working with the variable
$\sigma=(t-t_1)/|Q|=-1-P/|Q|$. The first form of the quartic is
\begin{eqnarray}
(B-2)\sigma^4&+&2[(B-2)(2+\cos\theta)-(p_0+\cos\theta)]\sigma^3\nonumber \\
	              &+&2(1+\cos\theta)[3(B-p_0-3)\sigma^2
					     +2(B-3p_0-5)\sigma-4(1+p_0)]=0,
\end{eqnarray}
where $p_0=P_0/|Q|>1$. Let $c_n$ denotes the coefficient of
$\sigma^n$.  Each coefficient is manifestly either zero or negative
for $B \leq 2$, and hence there can then be no roots with $\sigma>0$
i.e., $t>t_1$.  For $B>2$, $c_4>0$, $c_0<0$, and
$c_2-c_1=2(1+\cos\theta)(B+1+3p_0)>0$ and so $c_2>c_1$.  We now show
that, regardless of the sign of $c_2$, there can be only one sign
change from positive to negative in the sequence of coefficients
$(c_4,c_3,c_2,c_1,c_0)$.  If $c_2\geq 0$, then $B\geq p_0+3$ and
$c_3\geq 4+2p_0(1+\cos\theta)>0$, and the single sign change occurs
somewhere between $c_2$ and $c_0$. If $c_2<0$, there is a single sign
change somewhere between $c_4$ and $c_2$, depending on the sign of
$c_3$. By Descartes' rule of signs, the single sign change in the
sequence of coefficients means that the quartic has exactly one root
in $t_1>1$.

To show that there is one and only one root of the quartic in
$(t_1,t_2)$, we use a bilinear transformation to the variable
$\tau=(t-t_2)/(t_1-t)=(|Q|-P)/(|Q|+P)$ which maps the interval
$t_2<t<t_1$ to the whole of $\tau>0$.
Equation (A2) remains a quartic, and becomes
\begin{eqnarray}
         (1+\cos\theta)[(1+p_0)\tau^4+(B+p_0-1)\tau^3]
         -(1-\cos\theta)[(1+p_0-B)\tau+(p_0-1)]=0.\quad({\rm A}4)\nonumber
\end{eqnarray}
Now $c_4>0$, $c_3>0$, $c_2=0$, and $c_0<0$, when $c_n$ now
denotes the coefficient of $\tau^n$. There is a
single sign change from positive to negative
in the coefficient sequence regardless of the sign of $c_1$, and hence,
by Descartes' rule of signs, always one root for $t$ in $(t_2,t_1)$.
This root gives a point on the tangential caustic.

\setcounter{equation}{0}
\section{Third Order Moments}
Witt \& Mao (2000) give the third order moments in the case of on-axis
shear only.  The third order moments, derived from the residues of
simple poles at $t=0$, and double poles at $t=t_1$ and $t=t_2$ are
\begin{eqnarray}
	 \sum\mu_ip_ix^3_i
	 &=&{1\over 4}(R_1+3R_2)+
	 {B[(1-\gamma_1)\xi-\gamma_2\eta]^3
	    \over(1-\gamma^2_1-\gamma^2_2)^4},\\
	 \sum\mu_ip_ix^2_iy
	 &=&{q\over 4}(I_1+I_2)+
	 {B[(1-\gamma_1)\xi-\gamma_2\eta]^2
	    [(1+\gamma_1)\eta-\gamma_2\xi]
		 \over(1-\gamma^2_1-\gamma^2_2)^4},\\
	 \sum\mu_ip_ix_iy^2_i
	 &=&{q^2\over 4}(R_2-R_1)+
	 {B[(1-\gamma_1)\xi-\gamma_2\eta][(1+\gamma_1)\eta-\gamma_2\xi]^2
	    \over(1-\gamma^2_1-\gamma^2_2)^4},\\
	 \sum\mu_ip_iy^3_i
	 &=&{q^3\over 4}(3I_2-I_1)+
	 {B[(1+\gamma_1)\eta-\gamma_2\xi]^3
	    \over(1-\gamma^2_1-\gamma^2_2)^4}.
\end{eqnarray}
The $R_1$, $R_2$, $I_1$, and $I_2$ terms are all linear in the source
coordinates. They are the real and imaginary terms of the $m=3$,
$n=0$ and $m=2$, $n=1$ configuration moments respectively, and are
\begin{eqnarray}
	 R_1&=&{q^2B\over 4|Q|^3}
	 \left\{
	    \left\{
	       {1\over t^{B+1}_1}\left[1-{(B+1)|Q|\over t_1}\right]
	       -{1\over t^{B+1}_2}\left[1+{(B+1)|Q|\over t_2}\right]
	    \right\}
	 [\Re (Q)\xi-\gamma_2q^2\eta]
	 \right.\nonumber\\
	 &&\left.
	      -\left\{
	         {1\over t^{B+1}_1}\left[2+{(B+1)|Q|\over t_1}\right]
	         +{1\over t^{B+1}_2}\left[2-{(B+1)|Q|\over t_2}\right]
	      \right\}\right.\nonumber\\
           &&\left. \times
	   {[(\Re (Q))^2\xi+2\gamma_2q^2\Re(Q)\eta-\gamma^2_2q^2\xi]
		   \over |Q|}
		\right\},\nonumber\\
\noalign{\vskip7pt}
	 R_2&=&{q^2B\over 4|Q|^2}
	 \left\{
	    (B+1)|Q|\xi\left({1\over t^{B+2}_2}-{1\over t^{B+2}_1}\right)
	 \right.\nonumber\\
	 &&\left.
		 -\left[
		    (B+1)\left({1\over t^{B+2}_2}+{1\over t^{B+2}_1}\right)
			 +{1\over |Q|}\left({1\over 
t^{B+1}_1}-{1\over t^{B+1}_2}\right)
		  \right]
	    [\Re (Q)\xi+\gamma_2q^2\eta]
	 \right\},\nonumber\\
\noalign{\vskip7pt}
     I_1&=&{q^3B\over 4|Q|^3}
	 \left\{
	    \left\{
	       {1\over t^{B+1}_1}\left[1-{(B+1)|Q|\over t_1}\right]
	       -{1\over t^{B+1}_2}\left[1+{(B+1)|Q|\over t_2}\right]
	    \right\}
	 [\Re (Q)\eta+\gamma_2\xi]
	 \right.\nonumber\\
	 &&\left.
	      +\left\{
	         {1\over t^{B+1}_1}\left[2+{(B+1)|Q|\over t_1}\right]
	         +{1\over t^{B+1}_2}\left[2-{(B+1)|Q|\over t_2}\right]
	      \right\} \right.\nonumber\\
          &&\left.
	   {[(\Re (Q))^2\eta-2\gamma_2\Re (Q)\xi-\gamma^2_2q^2\eta]
		   \over |Q|}
		\right\},\nonumber\\
\noalign{\vskip7pt}
	 I_2&=&{q^3B\over 4|Q|^2}
	 \left\{
	    (B+1)|Q|\eta\left({1\over t^{B+2}_2}-{1\over t^{B+2}_1}\right)
	 \right.\nonumber\\
	 &&\left.
		 +\left[
		    (B+1)\left({1\over t^{B+2}_2}+{1\over t^{B+2}_1}\right)
			 +{1\over |Q|}\left({1\over 
t^{B+1}_1}-{1\over t^{B+1}_2}\right)
		  \right]
	    [\Re (Q)\eta-\gamma_2\xi]
	 \right\}.
\end{eqnarray}

\setcounter{equation}{0}
\section{Formulae for Hypergeometric Functions}
The hypergeometric functions that occur in this paper are all of the
form $_2F_1(M,jB;M+N;1-z)$, where $M$ and $N$ are integers and both
are positive. The following formula expresses them as two finite
sums, one with $N$ terms and the other with $M$ terms. It is obtained
by applying first Abramowitz and Stegun (1965, hereafter AS) equation
(15.3.9) and then their equation (15.3.5). The result is 
\begin{eqnarray}
\label{eq:hypergeom}
{(1\!-\!z)^{M\!+\!N\!-\!1} \over (M\!+\!N\!-\!1)!}
{}_2F_1(M,jB;M\!+\!N;1\!-\!z)
&=&{(-1)^M\over(N\!-\!1)!}\sum^{N\!-\!1}_{k=0} {{N\!-\!1\choose
k}(-z)^k\over\prod^{M\!+\!N\!-\!k\!-\!1}_{s=N\!-\!k} (jB\!-\!s)}\nonumber\\
&& +{z^{N\!-\!jB}\over(M\!-\!1)!}\sum^{M\!-\!1}_{k=0} {{M\!-\!1\choose
k}(-z)^k\over\prod^{N\!+\!k}_{s=1\!+\!k}(jB\!-\!s)}. 
\end{eqnarray}
It is significant that the products of terms in $B$ in the
denominators are all subparts of a longer product of terms in $B$
which multiplies the hypergeometric functions in the definition
(\ref{eq:eyethree}) of the coefficients $I_{j,\ell,N}$ which appear
in our series for the lensing invariants. Hence, these coefficients
never have poles in $B$ and are all analytic functions of $B$. For
example, the two $N=0$ integrals needed for the evaluation of the
extra term in equation (\ref{eq:fourmagval}) are:
\begin{eqnarray}
\label{eq:hypergeomtwo}
I_{1,2}&=&{t^{B\!-\!1}_1\over 16|Q|^3}
\left[\!-\!(B\!-\!2)(B\!-\!3)\!+\!2(B\!-\!1)(B\!-\!3)\left({t_2\over
t_1}\right)\!-\!(B\!-\!1)(B\!-\!2)\left({t_2\over t_1}\right)^2
\!+\!2\left({t_2\over t_1}\right)^{B\!-\!1}\right],\nonumber\\
I_{1,0}&=&{t^2_1t^{B\!-\!3}_2\over 16|Q|^3}
\left[(B\!-\!1)(B\!-\!2)\!-\!2(B\!-\!1)(B\!-\!3)\left({t_2\over
t_1}\right)\!+\!(B\!-\!2)(B\!-\!3)\left({t_2\over t_1}\right)^2
\!-\!2\left({t_2\over t_1}\right)^{3\!-\!B}\right]. \nonumber \end{eqnarray}
The similarity of these two expressions is an instance of the general
symmetry relation:
\begin{equation}
I_{j,\ell,N}(t_1,t_2)=I_{j,2j-\ell,N}(t_2,t_1). 
\end{equation}
Hence, only two distinct $I$ terms are needed for the first order
corrections for the first moments
\begin{eqnarray}
\sum_{4\ \images} \mu_i p_i x_i &=& {B( (1\!-\!\gamma_1)\xi\!-\!
\gamma_2 \eta) \over
(1\!-\!\gamma_1^2\!-\!\gamma_2^2)^2}\!+\!K^{10}_{30}
\xi^3\!+\!K^{10}_{21}
\xi^2 \eta\!+\! K^{10}_{12}\xi \eta^2\!+\! K^{10}_{03} \eta^3
\!+\!O(|\zeta|^5), \\
\sum_{4\ \images} \mu_i p_i y_i &=& {B(
(1\!+\!\gamma_1)\eta\!-\!\gamma_2 \xi) \over
(1\!-\!\gamma_1^2\!-\!\gamma_2^2)^2}\!+\!K^{01}_{30}
\xi^3\!+\!K^{01}_{21}
\xi^2 \eta \!+\! K^{01}_{12}\xi \eta^2\!+\!K^{01}_{03} \eta^3
\!+\!O(|\zeta|^5),
\end{eqnarray}
where
\begin{eqnarray}
K^{10}_{30}&=&b^2I_{1,0,1}\!+\!abI_{1,1,1}\!+\!abI_{1,2,1}\!+\!a^2I_{1,3,1},
\nonumber \\
K^{10}_{21}&=&cq[\!-\!3bI_{1,0,1}\!+\!(b\!-\!2a)I_{1,1,1}\!+\!(2b\!-\!a)
I_{1,2,1}\!+\!3aI_{1,3,1}],\nonumber \\
K^{10}_{12}&=&q^2[(ab\!+\!2c^2)I_{1,0,1}\!+\!(a^2\!-\!2c^2)I_{1,1,1}
\!+\!(b^2\!-\!2c^2)I_{1,2,1}\!+\!(ab\!+\!2c^2)I_{1,3,1}], \nonumber \\
K^{10}_{03}&=&cq^3[\!-\!aI_{1,0,1}\!+\!aI_{1,1,1}\!-\!bI_{1,2,1}\!+\!b
I_{1,3,1}], \nonumber \\
K^{01}_{03}&=&cq[\!-\!bI_{1,0,1}\!+\!bI_{1,1,1}\!-\!aI_{1,2,1}\!+\!a
I_{1,3,1}], \nonumber \\
K^{01}_{21}&=&q^2[(ab\!+\!2c^2)I_{1,0,1}\!+\!(b^2\!-\!2c^2)I_{1,1,1}\!+\!
(a^2\!-\!2c^2)I_{1,2,1}\!+\!(ab\!+\!2c^2)I_{1,3,1}], \nonumber \\
K^{10}_{12}&=&cq^3[\!-\!3aI_{1,0,1}\!+\!(a\!-\!2b)I_{1,1,1}\!+\!(2a\!-\!b)
I_{1,2,1}\!+\!3bI_{1,3,1}],\nonumber \\
K^{10}_{30}&=&q^4[a^2I_{1,0,1}\!+\!abI_{1,1,1}\!+\!abI_{1,2,1}\!+\!
b^2I_{1,3,1}],
\end{eqnarray}
and the constants $a$, $b$, and $c$ are
\begin{equation}
a={Bq^2 \over 2}\Biggl[ 1\!+\!{\Re\,Q\over 2|Q|}\Biggr], \quad
b={Bq^2 \over 2}\Biggl[ 1\!-\!{\Re\,Q\over 2|Q|}\Biggr], \quad
c={\gamma_2Bq^3 \over 2|Q|}.
\end{equation}
Finally, let us caution against using eq. (\ref{eq:hypergeom}) for
numerical calculations for all but small values of $M$ and $N$, and
for $z$ close to 1, because it can become a difference between two
nearly equal quantities. The hypergeometric series, which converges
rapidly for $z$ close to 1, should be used instead.
\setcounter{equation}{0}
\section{Convergence of Series and Analytic Continuation}
To prove that the series of equation (\ref{eq:fourmag}) converges for
sufficiently small $|\zeta|$, we need a bound on the coefficients
$I_{j,2m}$. The cases $B\!<\!2$ and $B\!>\!2$ require different
treatments. For $B\!<\!2$, we use the integral (\ref{eq:eyetwoint}),
replace the powers of $(w\!+\!t_1)$ in the denominator with the
smaller $(w\!+\!t_2)$, and derive the bound 
\begin{equation}
	 \left|I_{j,2m}\right|<{1\over\pi}\int^{\infty}_0
{w^{jB\!-\!1}dw\over(w\!+\!t_2)^{2j\!+\!2}}
=t^{(B\!-\!2)j-2}_2\int^{\infty}_0
	 {v^{jB\!-\!1}dv\over(v\!+\!1)^{2j\!+\!2}}. \end{equation}
The simpler integral can be evaluated in terms of Gamma functions
(AS, eq. 6.2.1). As this bound is independent of $m$, we can take it
out of the inner summation over $m$ when we use it in the infinite
series in equation (\ref{eq:fourmag}). The inner sum can then be
carried out to give simply
$(|\lambda|^2\!+\!|\nu|^2)^j=|\zeta|^{2j}$. The series for the
magnification is therefore majorized by the series 
\begin{equation}
{Bq^2\over\pi}\sum^{\infty}_{j=1}
{\Gamma[jB]\Gamma[2\!+\!(2\!-\!B)j]\over(2j\!+\!1)!}
t^{j(B\!-\!2)-2}_2|\zeta|^{2j}.
\end{equation}
Applying the ratio test and using Stirling's formula (AS, eq. 6.1.39)
to approximate the Gamma functions, we find that the series
(\ref{eq:fourmag}) converges if
\begin{equation}
\label{eq:extra}
|\zeta|<2\left({t_2\over 2\!-\!B}\right)^{(2\!-\!B)/2}B^{-B/2}.
\end{equation}
This condition is met uniformly for all $B$ in $[1,2]$ when 
\begin{equation}
\label{eq:zetaone}
|\zeta|<\sqrt{t_2},
\end{equation}
and then the infinite series (\ref{eq:fourmag}) for ${\cal M}(B)$ is
absolutely and uniformly convergent.

The integral (\ref{eq:eyetwoint}) is not valid for $B\!>\!2$, and we
must then use equation (\ref{eq:eyetwo}) and find a bound for the
hypergeometric function. We can use the hypergeometric series (AS,
eq. 15.1.1) for that function because its fourth argument lies in
$(0,1)$. As all the other arguments are positive, all the terms of
the infinite series are positive. Moreover, each of them is less than
the corresponding term of the infinite series for
$_2F_1(2j\!+\!2,jB;2j\!+\!2;1\!-\!t_2/t_1)$ because $\ell\leq 2j$.
This majorizing series is geometric (AS, eq. 15.1.8) and sums to
$(t_2/t_1)^{-jB}$. The terms $(jB\!-\!s)$ in the product are now all
positive for large enough $j$ because $s\leq 2j\!+\!1$, so we ignore
the $s$ terms and obtain, for large $j$, the bound 
\begin{equation}
	 \left|I_{j,2m}\right|<{(jB)^{2j\!+\!1}\over(2j\!+\!1)!}
t^{2m\!-\!2j\!-\!1}_2t^{jB\!-\!2m\!-\!1}_1. 
\end{equation}
The series for the magnification is now majorized by the series %
\begin{equation}
	 {Bq^2\over t_1t_2}\sum^{\infty}_{j=1}
	 {(jB)^{2j\!+\!1}t^{jB}_1\over(2j\!+\!1)!}\sum^{\infty}_{m=0}
{j\choose m}\left({|\lambda|\over t_1}\right)^{2m}
\left({|\nu|\over t_2}\right)^{2m\!-\!2j}. \end{equation}
The inner sum is
\begin{equation}
	 \left[{|\lambda|^2\over t^2_1}\!+\!{|\nu|^2\over t^2_2}\right]^j,
\end{equation}
and its maximum possible value is $|\zeta|^{2j}/t^{2j}_2$ because
$|\lambda|^2\!+\!|\nu|^2=|\zeta|^2$. Hence, the infinite series
(\ref{eq:fourmag}) for the magnification invariant is majorized for
$B>2$ by the series
\begin{equation}
	 {Bq^2\over t_1t_2}\sum^{\infty}_{j=1}
	 {(jB)^{2j\!+\!1}t^{jB}_1\over(2j\!+\!1)!} 	 \left({|\zeta|^2\over
t^2_2}\right)^j. \end{equation}
The ratio test show that this series is convergent, and that the
infinite series for the magnification is therefore absolutely
convergent, if
\begin{equation}
\label{eq:zetatwo}
	 |\zeta|<{2t_2\over eBt^{B/2}_1}.
\end{equation}
Thus, however large $B$ is, the series for the magnification
invariant converges for sufficiently small offset. Actually, if
$t_1$, the larger root of the equation $P^2=|Q|^2$, is less than
unity (as it is if the on-axis shear $\gamma_1$ is negative) and %
\begin{equation}
	 -\gamma_1>{(\gamma^2_1\!+\!\gamma^2_2)q^2\over 1-q^2},
\end{equation}
then
\begin{equation}
Bt^{B/2}_1\leq\left[{e\over 2}\ln\left({1\over
t_1}\right)\right]^{-1} \end{equation}
for all $B$, and the magnification series (\ref{eq:fourmag}) has a
finite radius of convergence that is independent of $B$. When $t_1=1$
(as it is when there is no shear), then the bound (\ref{eq:zetatwo})
decreases as $1/B$.

Regardless, if we need the magnification invariant ${\cal M}(B)$ for
a model for which $B=B_{\max}>2$, then we can find a $\delta$,
independent of $B$, such that the series for ${\cal M}(B)$ converges
absolutely for $|\zeta|<\delta$ for $1\leq B\leq B_{\max}$. That is,
our series is uniformly convergent as a function of $B$. The explicit
expressions that we have for each term in the series
(\ref{eq:fourmag}), from equation (\ref{eq:eyetwo}) and equation
(\ref{eq:hypergeom}) of Appendix C, shows that each is composed of
polynomials and exponential functions of $B$. Hence, each term of the
uniformly convergent series is an analytic function of $B$. It
follows, by a well-known theorem of complex analysis (e.g Levinson \&
Redheffer 1970), that the sum ${\cal M}(B)$ of the series is also an
analytic function of $B$ for $|\zeta|<\delta$. It can therefore be
continued beyond the range $1<B<2$ for which it was originally
derived by the analysis of \S 4.1 to apply to models with larger $B$.

With only minor changes, the series for the configuration invariants
can be handled in a similar way with a similar outcome; the infinite
series (\ref{eq:Smnseries}) for $S_{m,n}(B)$ is absolutely convergent
when the same inequalities (\ref{eq:zetaone}) and (\ref{eq:zetatwo})
as for ${\cal M}(B)$ hold. Hence, the series for the configuration
invariants of the four bright images also remain valid for $B>2$ for
sufficiently small offset $|\zeta|$.

Formulas for one of the four bright images become singular at the
radial caustic where it merges with the weak fifth image. Therefore we
can not expect our series to converge there, and inequality
(\ref{eq:zetatwo}) does not hold at the known analytical $|\zeta|$
values given in Appendix A for special points on the radial
caustic. However, the tangential caustic need not limit the
convergence, even though series lose physical significance there,
because the singularities of the two merging images can cancel. One
can verify that the $B<2$ inequality (\ref{eq:extra}) may indeed hold
at the cusps of the tangential caustic.

\end{appendix}

\end{document}